\documentclass[useAMS,usenatbib]{mn2e}
\usepackage{graphicx}
\usepackage{amssymb}
\usepackage{rotating}
\usepackage{times}
\title[The rest-frame $J$- and $H$-band luminosity functions to $z=3.5$]{The evolution of the rest-frame $J$- and $H$-band luminosity function of galaxies to $z=3.5$}
\author[M. Stefanon \& D. Marchesini]{Mauro Stefanon$^{1}$\thanks{E-mail:
mauro.stefanon@uv.es} and Danilo Marchesini$^{2}$\\ 
$^{1}$Observatori Astron\`omic Universitat de Val\`encia, C/ Catedr\'atico Agust\'in Escardino Benlloch, 7, 46980, Valencia, Spain \\
$^{2}$Tufts University, Robinson Hall, 212 College Avenue, Medford, MA 02155, USA\\}
\begin{document}

\date{Accepted --. Received --; in original form --}

\pagerange{\pageref{firstpage}--\pageref{lastpage}} \pubyear{2011}

\maketitle

\label{firstpage}

\begin{abstract}
We present the rest-frame $J$- and $H$- band luminosity function (LF) of field galaxies, based on a deep multi-wavelength composite sample from the MUSYC, FIRES and FIREWORKS survey public catalogues, covering a total area of 450 arcmin$^2$. The availability of flux measurements in the Spitzer IRAC  3.6, 4.5, 5.8, and 8 $\mu$m channels allows us to compute absolute magnitudes in the rest-frame $J$ and $H$ bands up to $z=3.5$ minimizing the dependence on the stellar evolution models.
We compute the LF in the four redshift bins $1.5<z<2.0$, $2.0<z<2.5$, $2.5<z<3.0$ and $3.0<z<3.5$. Combining our results with those already available at lower redshifts, we find that (1) the faint end slope is consistent with being constant up to $z=3.5$, with $\alpha=-1.05\pm0.03$ for the rest-frame $J$ band and $\alpha=-1.15\pm0.02$ for the rest-frame $H$ band; (2) the normalization $\phi^*$ decreases by a factor of 4-6 between $z=0$ and $z\simeq 1.75$ and by a factor of 2-3 between $z\simeq 1.75$ and $z=3.25$; (3) the characteristic magnitude $M^*$ shows a brightening from $z=0$ to $z\simeq 2$ followed by a slow dimming to $z=3.25$.
We finally compute the luminosity density (LD) in both rest-frame $J$ and $H$ bands. The analysis of our results together with those available in the literature shows that the LD is approximately constant up to $z\approx 1$, and it then decreases by a factor of 6 up to $z=3.5$.

\end{abstract}


\section{Introduction}

In the current concordance model, galaxies are the result of continuous mergers of dark matter halos driving baryonic matter assembly. In the last decades, simulations on halo occupation models have been able to quite accurately plot the formation of dark matter clusters. However, big uncertainties still remain at the time of translating dark matter haloes to what can actually be detected with our telescopes.

To this respect, the luminosity function (LF) of galaxies, i.e. the number density of galaxies per unit flux, is an extremely powerful tool to study the galaxy population and its evolution with cosmic time. Specifically, the analysis of the LF at different rest-frame wavelengths can give us information on different aspects of our present view of the Universe. The UV-optical LF allows for the study of the content and the evolution of the star formation rates with cosmic time. On the other hand the near infra-red (NIR) LF, being less sensitive to the absorption by dust and dominated by the light of older stars, is a better estimator of the overall stellar mass assembly of galaxies and of its rate of growth with time, revealing itself as a good test-bench for halo models.

The local NIR LF is still not yet well determined. Although a number of measurements have been derived so far, there seems to be uncertainties especially for the faint end slope $\alpha$. Estimates of the slope $\alpha$ range from $\approx -0.8$ (\citealt{bell2003}; \citealt{eke2005}), to $\alpha \approx -1.2$ (\citealt{jones2006}), with a median value around -1 (\citealt{mobasher93}; \citealt{glazebrook1995}; \citealt{cowie96}; \citealt{gardner97} and  \citealt{szokoly98}; \citealt{kochanek2001};  \citealt{cole2001} and  \citealt{hill2010}).

At even larger redshift, the LF determinations (most of which are done in the rest-frame $K_S$ band) still suffer from significant uncertainties (\citealt{saracco2006}). The faint-end slope seems to be always compatible with $\alpha=-1$ (\citealt{drory2003}; \citealt{pozzetti2003}; \citealt{dahlen2005}; \citealt{saracco2006}; \citealt{cirasuolo2010}) although these measurements suffer from the large uncertainties given by the limits in the depth of the photometric catalogues available so far.
There seems to be a general consensus however that the NIR LF does not significantly evolve to $z \approx 1$ with respect to the local LF ( \citealt{cowie96}; \citealt{pozzetti2003}; \citealt{drory2003}; \citealt{feulner2003}; \citealt{dahlen2005}).  A  brightening of the characteristic magnitude is instead found around $z \approx 1.2-1.5$ together with a decrease of the normalization, decrease that is seen up to $z=3$ (\citealt{saracco2006}; \citealt{cirasuolo2010}).

In this paper we present the rest-frame $J$- and $H$-bands LFs and luminosity density (LD) of field galaxies, obtained from three deep photometric redshift surveys, namely MUSYC, FIRES and FIREWORKS, complemented by deep \emph{Spitzer} 3.6, 4.5, 5.8, and 8 $\mu$m data.
As discussed in e. g. \citet{berta2007}, the combination of the Planck spectral peak from low-mass stars, the minimum in the $H^-$ opacity in stellar atmospheres and the molecular absorptions in the spectra of cold stars produce a maximum for the emission in the rest-frame NIR portion of galaxy spectra located at $1.6\mu$m (the so called \emph{$1.6\mu$m bump}). Furthermore, the AGN light can contribute significantly to the rest-frame $K$ band. Specifically, the contribution from the dust torus of the AGN can be in the rest-frame $K$ band a factor of 10 larger than in the rest-frame $J$ band, and a factor of ~4 larger than in the rest-frame $H$ band (e.g., \citealt{polletta2008}). The adoption of the rest-frame $J$ and $H$ bands makes thus the measurement of the LFs and LDs less sensitive to potential dust-obscured AGN contamination compared to measurements of the LFs in the rest-frame $K_s$, yet allowing us to sample a wavelength range dominated by stellar emission and very little affected by obscuration by dust.
The combination of depth and wavelength coverage in the mid-IR out to 8$\mu$m allows us to directly probe the rest-frame $J$ and $H$ bands out to $z\simeq 3.5$, relying more on observational data rather than on stellar population models, which are still significantly uncertain in the rest-frame NIR, due to different implementations of the TP-AGB phase \citep{maraston2005,conroy2009}. The total surveyed area sums to 450 arcmin$^2$ with complete U-to-8$\mu$m coverage, reducing thus the effects of cosmic variance, which we estimate to give on average a 15-20\% contribution. 

This paper is organized as follows. In section 2 we present the data sets used for this work together with a description on how we recover photometric redshifts and we select galaxies from the full sample. Section 3 presents the three methods adopted to estimate the LF and its associated uncertainties. In section 4 we present our results, while our conclusions are summarized in section 5.

Throughout this work, the adopted cosmology is $\Omega_{\Lambda}=0.7$, $\Omega_m=0.3$ and $H_0=70$ Km/s/Mpc. All magnitudes are expressed in the AB system.

\section{The sample}

For this work we used a total of seven public $K_s$-selected catalogues coming from three different deep multi-wavelength galaxy surveys covering the range from the optical to the Spitzer IRAC 8 $\mu$m waveband: the MUlti-walelength Survey by Yale-Chile (MUSYC - \citealt{danilo2009}), the Faint InfraRed Extragalactic Survey (FIRES - \citealt{labbe2003}, \citealt{fs2006}) and the  GOODS Chandra Deep Field-South (FIREWORKS - \citealt{wuyts2008}). Although they have all been presented in \citet{danilo2009}, for readers' sake these surveys will be briefly described in the following sections.

\subsection{MUSYC}

The deep NIR MUSYC survey consists of four $10' \times 10'$ fields, namely, Hubble Deep Field-South 1 and 2 (HDFS-1, HDFS-2, hereafter), the SDSS-1030 field, and the CW-1255 field, observed with the Infrared Side Port Imager (ISPI) camera at the Cerro Tololo Inter-American Observatory (CTIO) Blanco 4 m telescope, for a total surveyed area of 430 arcmin$^2$. A complete description of the deep NIR MUSYC observations, reduction procedures, and the construction of the $K$-selected catalog with $U$-to-$K$ photometry is presented in \citet{quadri2007}.
Deep Spitzer-IRAC 3.6-8.0 $\mu$m imaging is also available for the four fields. The average total limiting magnitudes of the IRAC images are 24.5, 24.2, 22.4, and 22.3 (3$\sigma$, AB magnitude) in the 3.6, 4.5, 5.8, and 8.0 $\mu$m bands, respectively. 
The $K$-selected catalogs with IRAC photometry included is publicly available at http://www.astro.yale.edu/musyc. The SDSS-1030, CW-1255, HDFS-1, and HDFS-2 catalogs are $K_S$ band-limited multicolor source catalogs down to $K_{S,tot}$ = 23.6, 23.4, 23.7, and 23.2, for a total of 3273, 2445, 2996, and 2118 sources, over fields of 109, 105, 109, 106 arcmin$^2$, respectively. All four fields were exposed in 14 different bands, U, B, V, R, I, z, J, H, K, and the four IRAC channels. The SDSS-1030, CW-1255, HDFS-1, and HDFS-2 K-selected catalogs have 90\% completeness levels at $K_{S,90}$=23.2, 22.8, 23.0, and 22.7, respectively. The final catalogs used in the construction of the composite sample have 2825, 2197, 2266, and 1749 objects brighter than the 90\% completeness in the $K_S$ band, over an effective area of 98.2, 91.0, 97.6, and 85.9 arcmin$^2$, respectively, for a total of 9037 sources over 372.7 arcmin$^2$.

\subsection{FIRES}

FIRES consists of two fields, namely, the Hubble Deep Field-South proper (HDF-S) and the field around MS 1054-03, a foreground cluster at z = 0.83. A complete description of the FIRES observations, reduction procedures, and the construction of photometric catalogs is presented in detail in \citet{labbe2003} and \citet{fs2006} for HDF-S and MS 1054-03 (hereafter HDFS and MS-1054), respectively. Both $K_S$-selected catalogs were later augmented with Spitzer-IRAC data (\citealt{wuyts2007}; \citealt{toft2007}). The HDFS catalog has 833 sources down to $K_{S,tot}$=26.0 over an area of $2.5 \times 2.5$ arcmin$^2$. The MS-1054 catalog has 1858 sources down to $K_{S,tot}$=25.0 over an area of $5.5 \times 5.3$ arcmin$^2$. The HDFS field was exposed in the WFPC2 U300, B450, V606, I814 passbands, the ISAAC $J_S$, H, and $K_S$ bands, and the four IRAC channels. The MS-1054 $K_S$-selected catalog comprises FORS1 U, B, V, WFPC2 V606 and I814, ISAAC $J_S$, H, and $K_S$, and IRAC 3.6-8.0 $\mu$m photometry. The HDFS and MS-1054 catalogs have 90\% completeness levels at $K_{S,90}$=25.5 and 24.1, respectively. The final HDFS and MS-1054 catalogs used in the construction of the composite sample have 715 and 1547 objects brighter than the 90\% completeness in the $K_S$ band, over an effective area of 4.5 and 21.0 arcmin$^2$, respectively.

\begin{figure*}
\centering
\includegraphics[width=19cm]{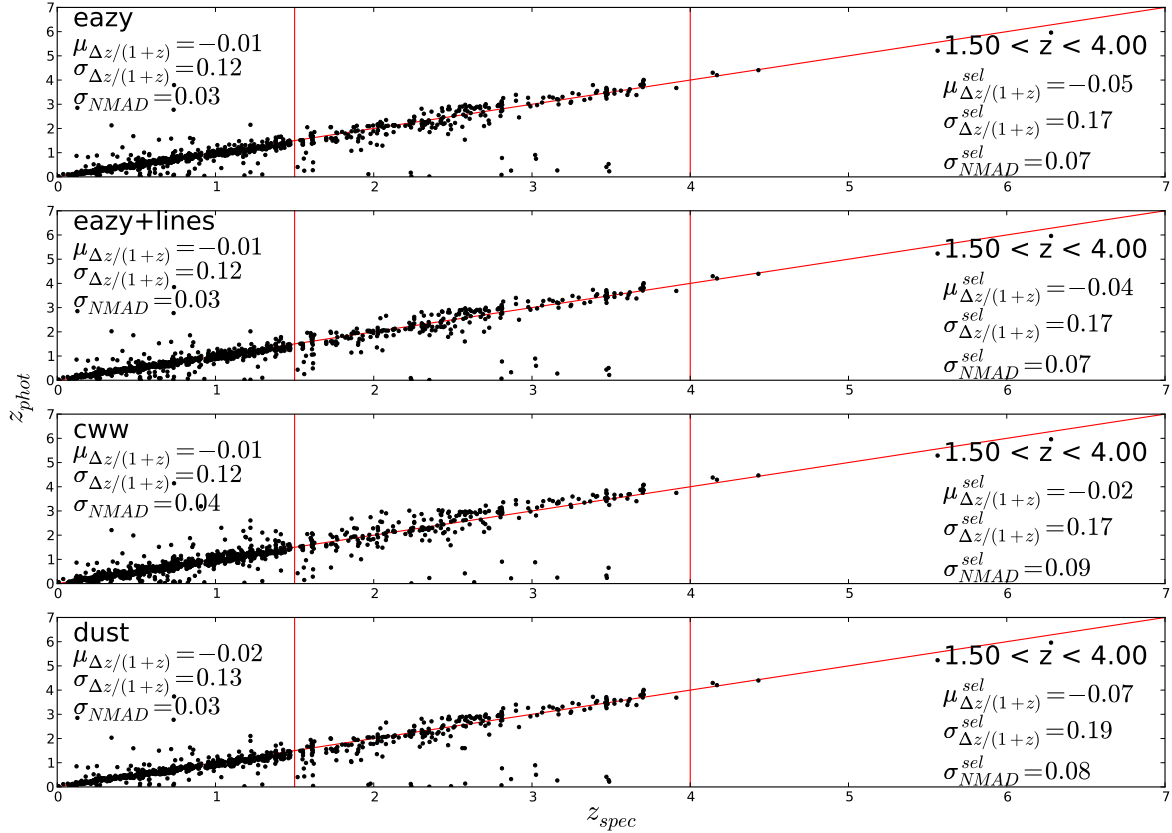}
\caption{Photometric redshifts compared to spectroscopic redshifts, for the four different SED sets. From top to bottom: EAZY default template set; EAZY default template set with the addition of Ly$\alpha$, H$\alpha$, H$\beta$, [OII] and [OIII]  emission lines; Coleman, Wu and Weedman template set and EAZY SED set with dusty galaxy template (see text for details). $\mu_{\rm \Delta z/(1+z)}$ and $\sigma_{\rm \Delta z/(1+z)}$ are the average and the standard deviation of $\Delta z/(1+z)$, respectively; $\sigma_{NMAD}$ is the normalized median absolute deviation.} 
\label{fig:phz_sed}
\end{figure*}

\subsection{FIREWORKS}

In this work, we adopted the $K_S$-selected catalog (dubbed FIREWORKS) of the CDFS field constructed based on the publicly available GOODS-CDFS data by \citet{wuyts2008}. The photometry was performed in an identical way to that of the FIRES fields, and the included passbands are the ACS B435, V606, i775, and z850 bands, the WFI U38, B, V, R, and I bands, the ISAAC J, H, and $K_S$ bands, and the four IRAC channels. The $K_S$-selected catalog comprises 6308 objects down to $K_{S,tot}$=24.6 over a total surveyed area of 138 arcmin$^2$; the variation in exposure time and observing conditions between the different ISAAC pointings lead to an inhomogeneous depth over the whole GOODS-CDFS field (hereafter CDFS). The final CDFS catalog used in the construction of the composite sample comprises 3559 objects brighter than the 90\% completeness level ($K_{S,90}$ = 23.7), over an effective area of 113 arcmin$^2$ with coverage in all bands.

\subsection{Photometric redshift and star/galaxy separation}

The downloaded catalogues all come with photometric redshift information; spectroscopic redshifts are also available for a small fraction of galaxies (around 10\% of the whole sample). However, we re-computed photometric redshifts, using the publicly available EAZY code (\citealt{brammer2008}), and adopting four different sets of Spectral Energy Distribution (SED) templates. The first SED set is the EAZY default template set; it consists of 5 SED templates built on the base of PEGASE models (\citealt{fr1997}), reproducing the colors of galaxies in the semi-analytic models by \citet{deLucia2007}, plus a template representing a 50 My galaxy with heavy dust obscuration. The second set is composed by a modified version of the standard EAZY templates, with the addition of Ly$\alpha$, H$\alpha$, H$\beta$, [OII] and [OIII] emission lines. The third is a set of six templates based on \citet{cww} colors, included in the Bayesian Photometric Redshift code (BPZ - \citealt{benitez2000}). The last set is an extension of the standard EAZY template set with the inclusion of a 1Gyr galaxy template, with $\tau=100$ Myr and $A_V=3$ mag, similar to the reddest template used in \citet{blanton2007}.

For all the four cases, the same default template error function and $K$-band prior was adopted.

Figure \ref{fig:phz_sed} shows the $z_{spec}$ vs. $z_{phot}$ plot for the four SED template sets used. The average $\Delta z/(1+z)$ are respectively -0.01, -0.01, -0.01, and -0.02 for the EAZY, EAZY+lines, CWW and EAZY+dust template sets when considering the full sample, and -0.05, -0.04, -0.02, and -0.07 when computed on the redshift range $1.5<z<4.0$. The standard deviations $\sigma_{\Delta z/(1+z)}$ are 0.12, 0.12, 0.12, and 0.13 respectively for the full sample and 0.17, 0.17, 0.17, and 0.19  in the redshift interval $1.5<z<4$. The normalized median absolute deviation, $\sigma_{\rm NMAD}$\footnote{The normalized median absolute deviation $\sigma_{\rm NMAD}$, defined as $1.48 \times median \big [ |(\Delta z - median(\Delta z))/(1 + z)| \big]$, is equal to the standard deviation for a Gaussian distribution, and it is less sensitive to outliers than the usual definition of the standard deviation (e.g., \citealt{ilbert2006}).}, is 0.03, 0.03, 0.04, and 0.03 over the whole redshift range, and 0.07, 0.07, 0.09, and 0.08 in the redshift interval $1.5<z<4$. The fraction of catastrophic photometric redshift ($n(\frac{\Delta z}{1+z})>5\sigma$) is 0.042, 0.046, 0.045 and 0.048 respectively.

We finally chose to adopt the standard EAZY template set, which is the one presenting the smallest deviation between spectroscopic and photometric redshifts in the targeted redshift range.

\begin{figure}
\includegraphics[width=8.5cm]{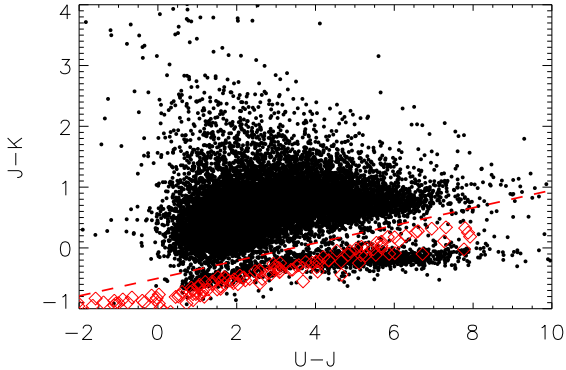}
\caption{Two colours diagram for all the sources in the composite catalogues. Colours from \citet{pickles1998} stellar atmosphere models are shown as red diamonds. The red dashed line represents Eq.~\ref{eq:star-gal}, which we adopted to separate stars from galaxies}
\label{fig:star_galaxy}
\end{figure}

The separation between stars and galaxy was done with the colour-colour diagram $(U-J)$ vs. $(J-K_s)$ (see figure \ref{fig:star_galaxy}). The same colours were also computed from the \citet{pickles1998} stellar atmosphere models, in order to improve the boundaries between stars and galaxy, especially for the reddest stars which fall out of the main sequence (see figure \ref{fig:star_galaxy} for full details). 

Galaxies were selected among the object satisfying the relation:
\begin{equation}
(J-K_s) \geq 0.145 \cdot (U-J)-0.45
\label{eq:star-gal}
\end{equation}

As a cross-check, we run the EAZY code with the \citet{pickles1998} model stellar atmosphere and checked that the objects identified as galaxies via the two-colour diagram had a $\chi^2$ greater than the $\chi^2$ obtained on the same object with the EAZY galaxy template set. This criteria was satisfied by all objects previously selected as galaxies with only 6.5\% of the objects selected as stars showing a discordant value for the $\chi^2$, giving confidence in our method to separate stars from galaxies.

A star-galaxy separation was also available in the original FIRES and FIREWORKS public catalogues. This selection was based on spectroscopy, SED-fitting with stellar templates and visual inspection of the object morphology \citep{rudnick2006,rudnick2003}. We verified that those objects selected as stars in the FIRES catalogue were actually falling in the correct region of our two-color plot.

The availability of \emph{Spitzer} IRAC data for all our sample allows us to compute absolute magnitudes in the rest-frame $J$ and $H$ bands with little dependance on the SED templates; in fact, as an extreme case, the rest-frame $H$ band at $z=3.5$ is shifted to the range $6.7-7.9\mu$m, well bracketed by the \emph{IRAC} channels 3 and 4, centered at 5.8 and 8$\mu$m.

Our final catalogue is composed by a total of 14295 galaxies, with redshifts determinations to $z=6.2$ and median redshift $z_{med}\simeq1$, distributed over an effective area of 450 arcmin$^2$. In the redshift range of interest, $1.5<z<3.5$, there is a total of 3496 objects, of which $\approx 6\%$ have spectroscopic redshifts.

\section{Methodology}
\label{methods}
For the measurement of the LF, we adopted three among the most widely used methods, namely the $1/V_{max}$ \citep{schmidt1968}, the STY maximum likelihood \citep{styml} and the Step-Wise Maximum Likelihood \citep{swml}.

The need to analyze composite samples with different photometric depth was overcome by applying standard techniques available for each chosen method. Below we summarize them.

\subsection{$1/V{max}$}

The $1/V_{max}$ method, first introduced by \citet{schmidt1968}, was then generalized by \citet{ab1980} to allow the simultaneous analysis of composite samples. This method presents a number of advantages: it is simple to code, it directly estimates the normalization of the LF, and it does not make any assumption on the spatial distribution of galaxies. The draw-back is that it is sensitive to the presence of clustering in the sample, affecting the estimate of the faint end.

In a given redshift interval $[z1,z2]$, the galaxy number density per unit magnitude in the $k$-th absolute magnitude bin $\Phi_k(M)$ is computed as follows:
\begin{equation}
\Phi_k(M)=\frac{1}{\Delta M}\sum_{i=1}^{N_{gal}}\frac{1}{V_{max,i}}
\end{equation}

where $\Delta M$ is the width of the magnitude bin and $N_{gal}$ is the number of galaxies in the redshift interval of interest. In the \emph{coherent} analysis proposed by \citet{ab1980}, $V_{max,i}$ is given by:
\begin{equation}
V_{max,i}=\sum_{j=1}^{n_s}\Omega_j \int_{z_{1}}^{min(z_{2,j},z_{max,i,j})} \frac{dV}{dz}dz
\end{equation}

with $n_s$ the number of samples constituting the full catalogue, $\Omega_j$ the apparent area in steradians corresponding the the $j$-th sample, $dV/dz$ the co-moving volume element per steradian and $z_{max,i,j}$ the maximum redshift at which the $i$-th galaxy could 
have been observed within the flux limit of the $j$-th sample. Standard Poisson errors were associated to each $\Phi_k(M)$.

\subsection{Step-wise maximum likelihood}

This method, developed by \citet{swml}, is based on a maximum likelihood estimate and with a non-parametric form.
This method relies on the fact that the number density of galaxies $\Phi(M, {\bf x} )$ can be factorialized into a function which depends on luminosity only, $\phi(M)$, and a factor depending on the position $f({\bf x})$ so that $\Phi(M,{\bf x})=\phi (M)\cdot f({\bf x})$. In particular, this means that the normalization factor of $\Phi$ is lost during the maximization of the likelihood, and should then be determined in some other way.
The LF $\Phi(M)$ is expressed as the sum over $n$ steps as:
\begin{equation}
\Phi(M)=\phi_i, M_i-\Delta M/2 <M < M_i+ \Delta M/2, i=1,..,n
\end{equation} 

By minimizing the natural logarithm of the likelihood expression, a recursive formula is found, allowing to compute the $\phi_i$. In order to take into account the non uniform magnitude limits of our data sets, we adopted the modification of the recursive expression as presented in \citet{hill2010}:

$$
\phi_i \Delta M= \frac{\sum_{k=1}^{N_{gal}}W(M_i-M_k)}{\sum_{k=1}^{N_{gal}} \left[ H(M_i-M_{f,k})/\sum_{j=1}^{N_{gal}}\phi_j\Delta M H(M_j-M_{f,k}) \right]}
$$
where $W$ is a \emph{window} function:
\begin{equation}
W(x)=\left\{
\begin{array}{rl}
1 & \mbox { if } -\Delta M/2 < x \leq \Delta M/2 \\
0 & \mbox{ otherwise}
\end{array}
\right.
\end{equation}
and
\begin{equation}
H(x)=\left\{
\begin{array}{rl}
1 & \mbox { if }  x \leq -\Delta M/2 \\
1/2 - x/ \Delta M  &  \mbox{ if }  -\Delta M/2 < x \leq \Delta M/2 \\
0 & \mbox{ otherwise}
\end{array}
\right.
\end{equation}

Uncertainties were computed by taking the square root of the first $n$ diagonal elements of the inverse of the information matrix (see \citealt{swml}).

The SWML method does not compute the normalization itself. For it, we rescaled the LF shape by applying the conservation of the number of galaxies and using the information from the $1/V_{max}$ method, as in \citet{hill2010}. Specifically, given an absolute magnitude range, let $N$ be the number of galaxies included in the given magnitude interval. The redshift edges allowing the selected sub-sample to be observed are then computed, together with the associated maximum co-moving volume $V_{max}$. The number density of galaxies obtained from the SWML estimate is then rescaled in order to match the $N/V_{max}$ value. The above method can thus be considered as a $1/V_{max}$ LF computed on a single wide absolute magnitude bin.

\subsection{STY maximum likelihood}
The third technique employed to compute the LF is the parametric method of \citet{styml}. 
The advantage is that the resulting LF is not binned, but instead it is a continuous function. On the other side, the shape of the LF is constrained by the model adopted. Similarly to the SWML, the STYML assumes that the LF can be separated into a term depending on  the spatial distribution and a term which depends on the luminosity distribution, thus losing the normalization. 

The expression for the probability $p_i$ of seeing galaxy $i$ in the sample can be written as:

\begin{equation}
p\propto \frac{\phi_i(M)}{\int_{M_f(m_s)}^{M_b(m_s)}\phi(M)dM}
\end{equation}
where the explicit dependence of the bright and faint absolute magnitude limits $M_b$ and $M_f$ to the apparent magnitude $m_s$ of the original catalogue allows to take into account the non uniform photometric depth across the seven catalogues constituting the final sample.

The $\phi(M)$ function was chosen to follow the \citet{schechter1976} distribution:
\begin{equation}
\phi(M)=(0.4\ln10)\phi^*10^{0.4(M^*-M)(1+\alpha)} \exp\left[-10^{0.4(M^*-M)}\right]
\end{equation}
where $\phi^*$ is the normalization, $\alpha$ is the faint-end slope, and $M^*$ is the characteristic magnitude indicating the change from the power-law to the exponential regime. The normalization factor $\phi^*$ was obtained by imposing the conservation of the number of galaxies in the total sample.

Confidence levels for $\alpha$ and $M^*$ corresponding to 68\%, 95\% and 99\% were computed from the ellipsoid of parameters defined by:
\begin{equation}
\ln \mathcal{L}=\ln \mathcal{L}_{max} - 0.5 \chi^2_{\beta}(N)
\label{eq:ellipsoid}
\end{equation}
where $\mathcal{L}$ is the maximum likelihood function and $\mathcal{L}_{max}$ its value at maximum, while $\chi^2_\beta(N)$ is the $\beta$-point of the $\chi^2$ distribution with $N$ degrees of freedom (\citealt{swml}). Uncertainties on the normalization factor $\phi^*$ were computed from the range of values compatible with the $1\sigma$ uncertainties in the $\alpha$ and $M^*$ parameters.

\subsection{Cosmic variance and photometric redshift uncertainties}

\begin{figure}
\includegraphics[width=9cm]{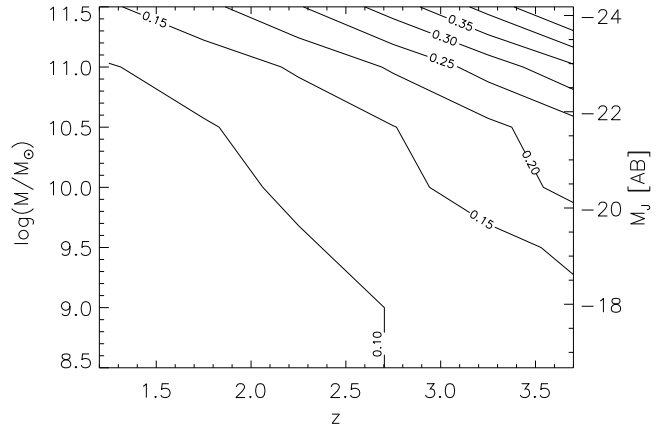}
\caption{Cosmic variance estimate for our survey geometry following \citet{moster2010}, as a function of redshift and mass (left vertical axis) and absolute magnitude (right vertical axis).}
\label{fig:cosvar}
\end{figure}

The data set used for our measurement of the LF is the combination of seven catalogues, each one related to a different region of the sky. This allows to keep in principle the effects of cosmic variance to low levels.

\begin{figure*}
\includegraphics{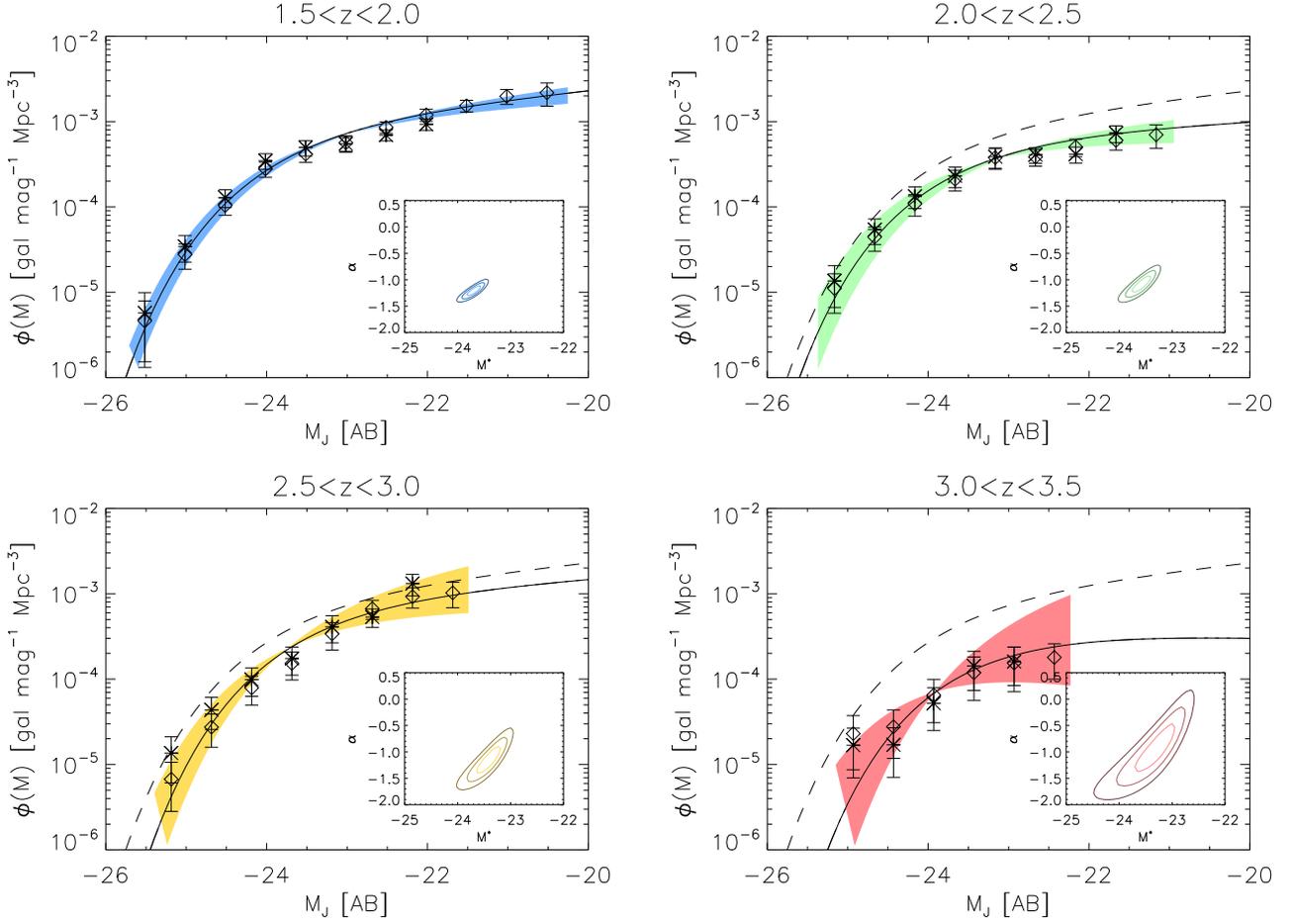}
\caption{Luminosity function for the rest-frame $J$ band, in the four redshift bins. Asterisks represent the $1/V_{max}$ measurement, diamonds are from the SWML. The Schechter function obtained via the maximum likelihood is shown as the solid black line. The coloured area in each plot indicates the 1 $\sigma$ (68\% confidence level) from the parametric maximum likelihood. The inset shows the 1, 2 and 3 $\sigma$ contours (corresponding to 68\%, 95\% and 98\% confidence region) for the joint $\alpha$-$M^*$ parameters from the ML analysis. The dashed line marks our $1.5<z<2.0$ LF.}
\label{fig:lf_j}
\end{figure*}

In this work, we included a more refined estimate of cosmic variance following the recipe by \citet{moster2010}. A halo distribution model is used to relate the stellar mass to the dark matter halo as a function of redshift; the galaxy bias is then estimated via dissipation-less N-body simulations. The cosmic variance is first computed on dark matter haloes, and then converted to galaxy cosmic variance by applying the galaxy bias. This estimate was cross-checked with the different evaluation of cosmic variance by \citet{driver2010}. Their work is based on direct computation of the cosmic variance using $M^*$ galaxies from the SDSS catalogue. The expression found is then generalized to any redshift bin amplitude and mean value and to any geometry of the survey. We find that the two estimates, in the case of $M^*$ galaxies, are consistent within 70\% in the lowest redshift bin, but differ up to a factor of 2.5 in the higher redshift ranges. As discussed in \citet{driver2010}, this discrepancy can be explained as the change in $M^*$ stellar mass value with redshift.

The computation of the luminosity as a function of mass (or, more frequently, the computation of mass from the luminosity), necessary to obtain the values for cosmic variance  is generally a non trivial task, involving the generation of synthetic SEDs based on different initial mass functions, which are then fitted on a per-galaxy basis. For our purposes of cosmic variance estimate in the final LF, we performed the conversion between galaxy baryonic mass $M$ and luminosity $L_{J_{AB}}$ and $L_{H_{AB}}$ a-posteriori on the LF, under the work hypothesis that the mass-to-light ratio can be considered constant over all the involved luminosity range and equal to its average value. We adopted the mean value $\langle M/L\rangle=1.0_{-0.27}^{+0.32}M_{\odot}/L_{\odot}$ from \citet{cole2001} for both the $J$ and $H$ bands.

\begin{figure*}
\includegraphics{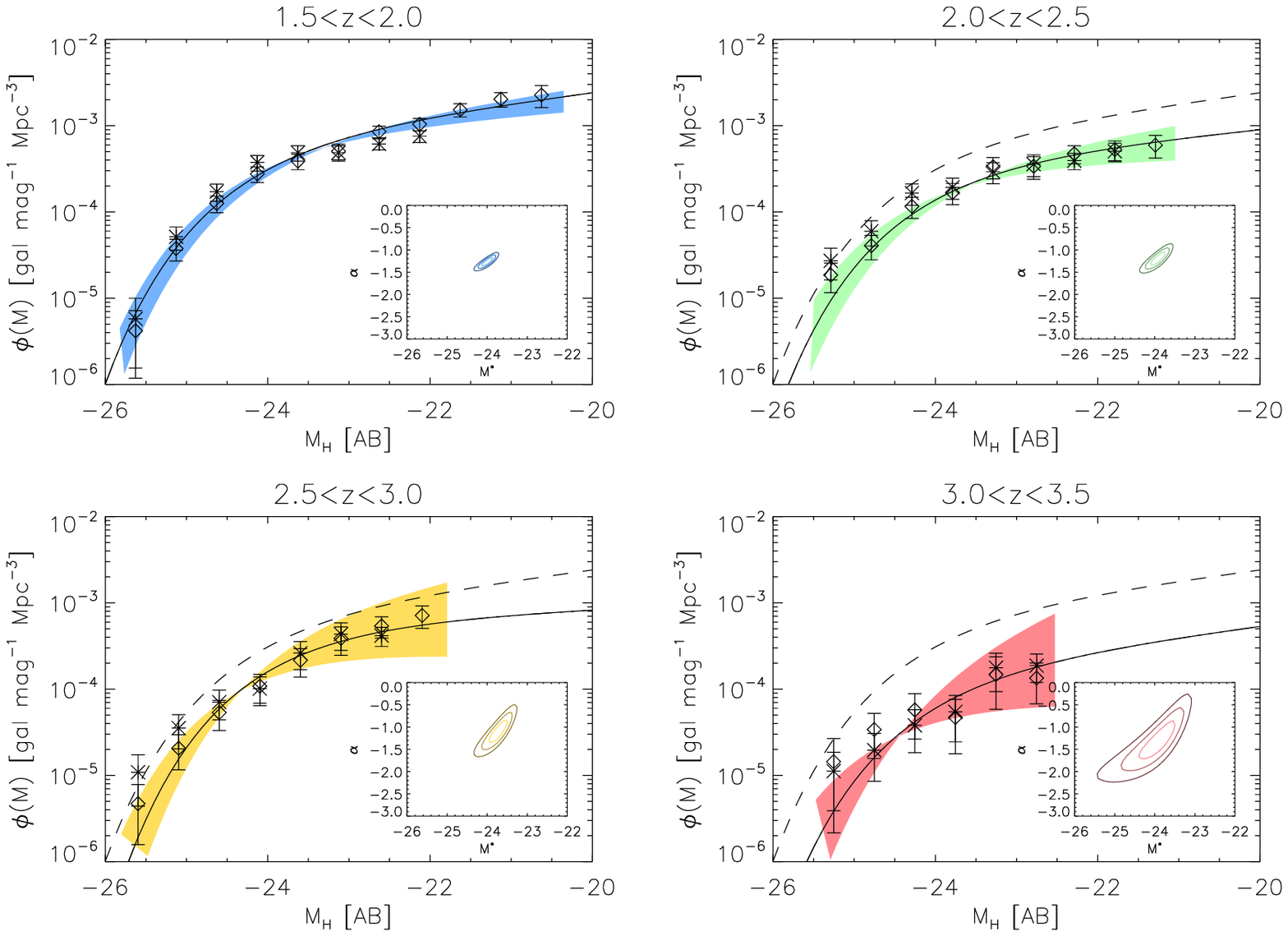}
\caption{Luminosity function in the H rest frame filter. Plot conventions are the same as for Figure \ref{fig:lf_j}.}
\label{fig:lf_h}
\end{figure*}

Figure \ref{fig:cosvar} shows as a contour plot the values of the cosmic variance for our data as a function of redshift and galaxy mass, as computed using the Moster et al. cookbook. The values of cosmic variance range from 0.07 to 0.19 in the lowest redshift bin, from 0.09 to 0.24 in the $2.0<z<2.5$ bin, from 0.10 to 0.32 in the $2.5<z<3.0$ bin and from 0.12 to 0.42 in the $3.0<z<3.5$ redshift bin. The recovered uncertainties have been added in quadrature to the standard errors computed in the $1/V_{\rm max}$ and SWML methods, while the cosmic variance corresponding to $M^*$ has been added in quadrature to the error on $\phi^*$.

The effects of photometric redshift errors have been studied via Monte Carlo simulations. Five hundred realizations of the LF in each redshift bin were  computed. The redshift of each source in the original catalogue was randomly modified according to the gaussian standard deviation recovered from figure \ref{fig:phz_sed}; the absolute magnitude of each object was then modified accordingly. The distribution of parameters of the recovered LF did not show any systematic effect and the spread of the parameters was compatible with the photometric redshift errors, consistent with the Monte Carlo simulations performed in \citet{marchesini2007}.

\section{$J$- and $H$-band Luminosity Functions}

\begin{table}
\begin{tabular}{ccccc}
z range & $\alpha$  & $M^* $ & $\phi^*$ ($10^{-4}$ Mag$^{-1}$ Mpc$^{-3}$) \\
\hline
1.5-2.0 & $-1.24^{+0.03}_{-0.03}$ & $-23.72^{+0.09}_{-0.06}$ & $11.31^{+0.28}_{-0.10}$ \\
2.0-2.5 & $-1.12^{+0.11}_{-0.13}$ & $-23.60^{+0.14}_{-0.17}$ & $7.45^{+0.53}_{-0.38}$ \\
2.5-3.0 & $-1.17^{+0.18}_{-0.22}$ & $-23.42^{+0.22}_{-0.23}$ & $9.73^{+5.37}_{-2.33}$ \\
3.0-3.5 & $-0.92^{+0.42}_{-0.48}$ & $-23.28^{+0.33}_{-0.39}$ & $4.36^{+8.61}_{-1.93}$
\end{tabular}
\caption{Schechter parameters for the J LF from the maximum likelihood analysis with one $\sigma$ errors, including uncertainties from cosmic variance.}
\label{tab:schechter_J}
\end{table}

\begin{table}
\begin{tabular}{ccccc}
z range & $\alpha$  & $M^* $ & $\phi^*$ ($10^{-4}$ Mag$^{-1}$ Mpc$^{-3}$) \\
\hline
1.5-2.0 & $-1.30^{+0.04}_{-0.03}$ & $-24.03^{+0.06}_{-0.05}$ & $8.79^{+0.50}_{-0.20}$ \\
2.0-2.5 & $-1.23^{+0.12}_{-0.07}$ & $-23.94^{+0.13}_{-0.12}$ & $4.35^{+0.68}_{-0.45}$ \\
2.5-3.0 & $-1.11^{+0.19}_{-0.18}$ & $-23.74^{+0.21}_{-0.23}$ & $6.32^{+4.52}_{-1.36}$ \\
3.0-3.5 & $-1.30^{+0.40}_{-0.49}$ & $-23.89^{+0.36}_{-0.42}$ & $2.03^{+8.50}_{-2.04}$
\end{tabular}
\caption{Schechter parameters for the H LF from the maximum likelihood analysis with one $\sigma$ errors, including uncertainties from cosmic variance.}
\label{tab:schechter_H}
\end{table}

\begin{table}
\resizebox{8.7cm}{!}{
\begin{tabular}{cccc}
$z$ range & $M_J [AB]$ & $SWML  [h_{70}^3 mag^{-1} Mpc^{-3}] $ & $1/V_{\rm max} [h_{70}^3 mag^{-1} Mpc^{-3}]$\\
\hline
$1.5<z<2.0$ & $ -25.51 $  & $4.78~(\pm 3.30) \pm 3.42 \times 10^{-6}$  &  $5.65 ~(\pm 3.99) \pm 4.13 \times 10^{-6}$  \\
&$ -25.01 $  & $2.87 ~(\pm 0.77) \pm 0.94 \times 10^{-5}$  &  $3.39 ~(\pm 0.98) \pm 1.16 \times 10^{-5}$  \\
&$ -24.51 $  & $1.08 ~(\pm 0.14) \pm 0.25 \times 10^{-4}$  &  $1.27 ~(\pm 0.19) \pm 0.30 \times 10^{-4}$  \\
&$ -24.01 $  & $2.90 ~(\pm 0.25) \pm 0.59 \times 10^{-4}$  &  $3.42 ~(\pm 0.31) \pm 0.71 \times 10^{-4}$  \\
&$ -23.51 $  & $4.44~ (\pm 0.35) \pm 0.90 \times 10^{-4}$  &  $5.03 ~(\pm 0.38) \pm 1.01 \times 10^{-4}$  \\
&$ -23.01 $  & $5.90 ~(\pm 0.49) \pm 1.20 \times 10^{-4}$  &  $5.82 ~(\pm 0.43) \pm 1.17 \times 10^{-4}$  \\
&$ -22.51 $  & $9.09 ~(\pm 0.84) \pm 1.38 \times 10^{-4}$  &  $7.80 ~(\pm 0.62) \pm 1.13 \times 10^{-4}$  \\
&$ -22.01 $  & $1.26 ~(\pm 0.14) \pm 0.20 \times 10^{-3}$  &  $9.39 ~(\pm 0.84) \pm 1.41 \times 10^{-4}$  \\
&$ -21.51 $  & $1.61 ~(\pm 0.20) \pm 0.25 \times 10^{-3}$  &  -  \\
&$ -21.01 $  & $2.08 ~(\pm 0.36) \pm 0.41 \times 10^{-3}$  &  -  \\
&$ -20.51 $  & $2.29 ~(\pm 0.66) \pm 0.69 \times 10^{-3}$  &  -  \\
&&&\\
$ 2.0<z<2.5$ & $ -25.17 $  & $1.36 (\pm 0.52) \pm 0.61 \times 10^{-5}$  &  $1.61 ~(\pm 0.66) \pm 0.77 \times 10^{-5}$  \\
&$ -24.67 $  & $4.52 ~(\pm 0.91) \pm 1.44 \times 10^{-5}$  &  $5.36 ~(\pm 1.20) \pm 1.78 \times 10^{-5}$  \\
&$ -24.17 $  & $1.15 ~(\pm 0.15) \pm 0.32 \times 10^{-4}$  &  $1.38 ~(\pm 0.19) \pm 0.39 \times 10^{-4}$  \\
&$ -23.67 $  & $2.24 ~(\pm 0.25) \pm 0.61 \times 10^{-4}$  &  $2.65 ~(\pm 0.29) \pm 0.71 \times 10^{-4}$  \\
&$ -23.17 $  & $4.08 ~(\pm 0.47) \pm 1.11 \times 10^{-4}$  &  $4.48 ~(\pm 0.48) \pm 1.20 \times 10^{-4}$  \\
&$ -22.67 $  & $4.22 ~(\pm 0.64) \pm 0.92 \times 10^{-4}$  &  $4.36 ~(\pm 0.58) \pm 0.89 \times 10^{-4}$  \\
&$ -22.17 $  & $5.52 ~(\pm 0.93) \pm 1.27 \times 10^{-4}$  &  $4.24 ~(\pm 0.59) \pm 0.89 \times 10^{-4}$  \\
&$ -21.67 $  & $6.51 ~(\pm 1.28) \pm 1.50 \times 10^{-4}$  &  $7.48 ~(\pm 1.41) \pm 1.67 \times 10^{-4}$  \\
&$ -21.17 $  & $7.58 ~(\pm 2.12) \pm 2.31 \times 10^{-4}$  &  -  \\
&&&\\
$2.5<z<3.0$ & $ -25.19 $  & $8.41 ~(\pm 3.73) \pm 4.62 \times 10^{-6}$  &  $1.61 ~(\pm 0.66) \pm 0.84 \times 10^{-5}$  \\
&$ -24.69 $  & $2.82 ~(\pm 0.76) \pm 1.19 \times 10^{-5}$  &  $4.43 ~(\pm 1.11) \pm 1.81 \times 10^{-5}$  \\
&$ -24.19 $  & $8.45 ~(\pm 1.54) \pm 3.14 \times 10^{-5}$  &  $1.21 ~(\pm 0.21) \pm 0.44 \times 10^{-4}$  \\
&$ -23.69 $  & $1.68 ~(\pm 0.27) \pm 0.60 \times 10^{-4}$  &  $2.19 ~(\pm 0.36) \pm 0.79 \times 10^{-4}$  \\
&$ -23.19 $  & $3.78 ~(\pm 0.53) \pm 1.33 \times 10^{-4}$  &  $4.35 ~(\pm 0.58) \pm 1.52 \times 10^{-4}$  \\
&$ -22.69 $  & $7.48 ~(\pm 1.07) \pm 1.88 \times 10^{-4}$  &  $5.80 ~(\pm 0.73) \pm 1.40 \times 10^{-4}$  \\
&$ -22.19 $  & $1.03 ~(\pm 0.17) \pm 0.27 \times 10^{-3}$  &  -  \\
&$ -21.69 $  & $1.14 ~(\pm 0.29) \pm 0.38 \times 10^{-3}$  &  -  \\
&&&\\
$3.0<z<3.5$ & $ -24.93 $  & $2.64 ~(\pm 1.25) \pm 1.67 \times 10^{-5}$  &  $1.83 ~(\pm 0.75) \pm 1.07 \times 10^{-5}$  \\
&$ -24.43 $  & $3.81 ~(\pm 1.42) \pm 2.13 \times 10^{-5}$  &  $2.51 ~(\pm 1.13) \pm 1.54 \times 10^{-5}$  \\
&$ -23.93 $  & $9.80 ~(\pm 1.67) \pm 4.43 \times 10^{-5}$  &  $7.45 ~(\pm 2.23) \pm 3.83 \times 10^{-5}$  \\
&$ -23.43 $  & $1.68 ~(\pm 0.14) \pm 0.72 \times 10^{-4}$  &  $1.54 ~(\pm 0.38) \pm 0.75 \times 10^{-4}$  \\
&$ -22.93 $  & $2.19 ~(\pm 0.25) \pm 0.95 \times 10^{-4}$  &  $2.82 ~(\pm 1.35) \pm 1.79 \times 10^{-4}$  \\
&$ -22.43 $  & $2.55 ~(\pm 0.47) \pm 0.83 \times 10^{-4}$  &  -  \\
\end{tabular}}
\caption{Luminosity function values obtained with the SWML and $1/V_{\rm max}$ methods in the four redshift bins for the $J$ filter. The first error value refers to the error estimated via the information matrix for the SWML method and to the Poisson error for the $1/V_{\rm max}$ method, while the second term is the cumulative error taking into account also cosmic variance uncertainties.}
\label{tab:lf_j}
\end{table}

\begin{table}
\resizebox{8.7cm}{!}{
\begin{tabular}{cccc}
$z$ range & $M_H [AB]$ & $SWML  [h_{70}^3 mag^{-1} Mpc^{-3}] $ & $1/V_{\rm max} [h_{70}^3 mag^{-1} Mpc^{-3}]$\\
\hline
$1.5<z<2.0$&$ -25.63 $  & $4.12 ~(\pm 2.85 \pm 2.95 \times 10^{-6}$  &  $5.69 ~(\pm 4.02) \pm 4.16 \times 10^{-6}$  \\
&$ -25.13 $  & $3.71 ~(\pm 0.79) \pm 1.05 \times 10^{-5}$  &  $5.08 ~(\pm 1.20) \pm 1.53 \times 10^{-5}$  \\
&$ -24.63 $  & $1.24 ~(\pm 0.14) \pm 0.27 \times 10^{-4}$  &  $1.69 ~(\pm 0.22) \pm 0.38 \times 10^{-4}$  \\
&$ -24.13 $  & $2.74 ~(\pm 0.23) \pm 0.56 \times 10^{-4}$  &  $3.73 ~(\pm 0.33) \pm 0.77 \times 10^{-4}$  \\
&$ -23.63 $  & $3.92 ~(\pm 0.32) \pm 0.80 \times 10^{-4}$  &  $4.91 ~(\pm 0.37) \pm 0.99 \times 10^{-4}$  \\
&$ -23.13 $  & $5.04 ~(\pm 0.44) \pm 1.04 \times 10^{-4}$  &  $4.86 ~(\pm 0.38) \pm 0.98 \times 10^{-4}$  \\
&$ -22.63 $  & $8.61 ~(\pm 0.81) \pm 1.32 \times 10^{-4}$  &  $6.19 ~(\pm 0.50) \pm 0.90 \times 10^{-4}$  \\
&$ -22.13 $  & $1.04 ~(\pm 0.12) \pm 0.17 \times 10^{-3}$  &  $7.49 ~(\pm 0.73) \pm 1.16 \times 10^{-4}$  \\
&$ -21.63 $  & $1.53 ~(\pm 0.19) \pm 0.27 \times 10^{-3}$  &  -  \\
&$ -21.13 $  & $2.03 ~(\pm 0.34) \pm 0.39 \times 10^{-3}$  &  -  \\
&$ -20.63 $  & $2.26 ~(\pm 0.61) \pm 0.64 \times 10^{-3}$  &  -  \\
&&&\\
$2.0<z<2.5$&$ -25.29 $  & $2.04 ~(\pm 0.55) \pm 0.74 \times 10^{-5}$  &  $2.95 ~(\pm 0.89) \pm 1.15 \times 10^{-5}$  \\
&$ -24.79 $  & $4.06 ~(\pm 0.80) \pm 1.28 \times 10^{-5}$  &  $5.89 ~(\pm 1.26) \pm 1.92 \times 10^{-5}$  \\
&$ -24.29 $  & $1.23 ~(\pm 0.15) \pm 0.34 \times 10^{-4}$  &  $1.74 ~(\pm 0.22) \pm 0.48 \times 10^{-4}$  \\
&$ -23.79 $  & $1.76 ~(\pm 0.21) \pm 0.48 \times 10^{-4}$  &  $2.02 ~(\pm 0.23) \pm 0.55 \times 10^{-4}$  \\
&$ -23.29 $  & $3.46 ~(\pm 0.44) \pm 0.96 \times 10^{-4}$  &  $2.97 ~(\pm 0.34) \pm 0.81 \times 10^{-4}$  \\
&$ -22.79 $  & $3.62 ~(\pm 0.57) \pm 1.06 \times 10^{-4}$  &  $3.75 ~(\pm 0.51) \pm 1.05 \times 10^{-4}$  \\
&$ -22.29 $  & $5.04 ~(\pm 0.88) \pm 1.18 \times 10^{-4}$  &  $4.36 ~(\pm 0.71) \pm 0.99 \times 10^{-4}$  \\
&$ -21.79 $  & $5.56 ~(\pm 1.14) \pm 1.44 \times 10^{-4}$  &  $5.05 ~(\pm 0.94) \pm 1.23 \times 10^{-4}$  \\
&$ -21.29 $  & $6.27 ~(\pm 1.69) \pm 1.84 \times 10^{-4}$  &  -  \\
&&&\\
$2.5<z<3.0$&$ -25.60 $  & $5.69 ~(\pm 2.93) \pm 3.46 \times 10^{-6}$  &  $1.34 ~(\pm 0.60) \pm 0.74 \times 10^{-5}$  \\
&$ -25.10 $  & $1.91 ~(\pm 0.58) \pm 0.84 \times 10^{-5}$  &  $3.49 ~(\pm 0.97) \pm 1.49 \times 10^{-5}$  \\
&$ -24.60 $  & $5.21 ~(\pm 1.06) \pm 1.99 \times 10^{-5}$  &  $6.98 ~(\pm 1.37) \pm 2.64 \times 10^{-5}$  \\
&$ -24.10 $  & $1.05 ~(\pm 0.18) \pm 0.39 \times 10^{-4}$  &  $9.98 ~(\pm 1.72) \pm 3.65 \times 10^{-5}$  \\
&$ -23.60 $  & $2.20 ~(\pm 0.35) \pm 0.79 \times 10^{-4}$  &  $2.67 ~(\pm 0.39) \pm 0.95 \times 10^{-4}$  \\
&$ -23.10 $  & $3.93 ~(\pm 0.65) \pm 1.43 \times 10^{-4}$  &  $4.40 ~(\pm 0.59) \pm 1.54 \times 10^{-4}$  \\
&$ -22.60 $  & $5.56 ~(\pm 1.03) \pm 1.54 \times 10^{-4}$  &  $4.25 ~(\pm 0.60) \pm 1.06 \times 10^{-4}$  \\
&$ -22.10 $  & $7.23 ~(\pm 1.51) \pm 2.12 \times 10^{-4}$  &  -  \\
&&&\\
$3.0<z<3.5$&$ -25.25 $  & $1.40 ~(\pm 1.02) \pm 1.17 \times 10^{-5}$  &  $1.10 ~(\pm 0.55) \pm 0.72 \times 10^{-5}$  \\
&$ -24.75 $  & $3.32 ~(\pm 1.13) \pm 1.79 \times 10^{-5}$  &  $1.93 ~(\pm 0.73) \pm 1.09 \times 10^{-5}$  \\
&$ -24.25 $  & $6.10 ~(\pm 1.98) \pm 3.23 \times 10^{-5}$  &  $4.11 ~(\pm 1.19) \pm 2.10 \times 10^{-5}$  \\
&$ -23.75 $  & $4.69 ~(\pm 2.10) \pm 2.87 \times 10^{-5}$  &  $5.38 ~(\pm 1.94) \pm 2.97 \times 10^{-5}$  \\
&$ -23.25 $  & $1.47 ~(\pm 0.62) \pm 0.87 \times 10^{-4}$  &  $1.77 ~(\pm 0.39) \pm 0.84 \times 10^{-4}$  \\
&$ -22.75 $  & $1.35 ~(\pm 0.54) \pm 0.65 \times 10^{-4}$  &  $1.88 ~(\pm 0.45) \pm 0.68 \times 10^{-4}$  \\
\end{tabular}}
\caption{Luminosity function values obtained with the SWML and $1/V_{\rm max}$ methods in the four redshift bins for the $H$ filter. The first error value refers to the error estimated via the information matrix for the SWML method and to the Poisson error for the $1/V_{\rm max}$ method, while the second term is the cumulative error taking into account also cosmic variance uncertainties.}
\label{tab:lf_h}
\end{table}

Both rest-frame $J$- and $H$-band LFs were estimated in the redshift intervals $1.5<z<2.0$, $2.0<z<2.5$, $2.5<z<3.0$, and $3.0<z<3.5$ with the three methods described in Sec.~\ref{methods}. In Table \ref{tab:lf_j} and Table \ref{tab:lf_h} we present our measurements obtained with the SWML and $1/V_{\rm max}$ method, while the derived Schechter parameters in each filter and redshift range are summarized in Table \ref{tab:schechter_J} and  Table \ref{tab:schechter_H}.

Figure \ref{fig:lf_j} shows the LF for the rest-frame J filter in the four redshift bins. The number of objects used to construct the LF in each redshift bin are respectively 996, 419, 298 and 103. The three methods return consistent measurements of the LFs.

In Figure \ref{fig:lf_h} we show our measurement of the LF obtained in the H filter, for the same four redshift bins as for the $J$-band LF. The number of objects used to compute the LF is 996, 419, 298 and 103 respectively for the  $1.5<z<2.0$, $2.0<z<2.5$, $2.5<z<3.0$ and $3.0<z<3.5$ intervals. To date, this is the first measurement of the rest-frame $H$-band LF in the interval $z\in [1.5, 3.5]$. 

\subsection{Comparison with previous works}
\begin{figure}
\includegraphics[width=8.5cm]{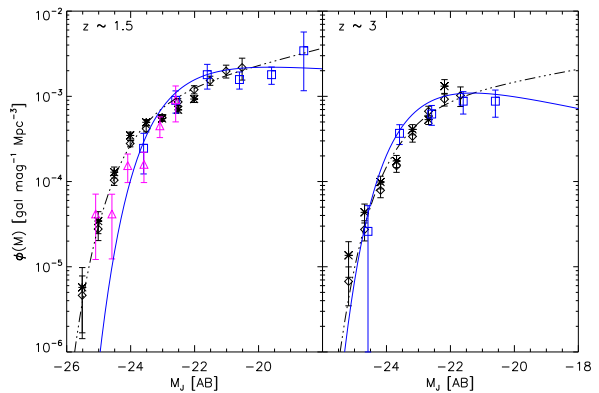}
\caption{Comparison of \citet{pozzetti2003} (magenta triangles - left panel) and \citet{saracco2006} LFs (blue squares and blue solid line) with our measurements (black symbols: asterisks for $1/V_{\rm max}$, diamonds for the SWML, dash-dotted line for the Schechter parameterization). The left panel refers to $z\approx 1.5$ LF, while the right panel to the $z \approx 3$ LF.}
\label{fig:comp_saracco}
\end{figure}

The measurements by \citet{pozzetti2003} and \citet{saracco2006}  are the only previously measured rest-frame $J$-band LFs computed in redshift ranges comparable with ours.

In \citet{pozzetti2003} the LF is computed for the rest-frame $J$ and rest-frame $K$ bands, using both spectroscopic and photometric data from the K20 survey \citep{cimatti2002}. Out of the three redshift ranges used to measure the LF ($z\in[0.2,0.65], [0.75,1.3], [1.3,1.9]$), the highest interval is the only one which we can compare to. Their rest-frame $J$-band LF is shown by the magenta triangles in the left panel of Fig. \ref{fig:comp_saracco}. Our data allow us to compute the LF down to $M_J=-20$, around two magnitudes fainter than their limit; in addition, the area covered by our sample is approximately 10 times larger than the area covered by the K20 survey, allowing us to estimate the bright end of the LF more robustly and to a 0.5 mag brighter limit. In the overlapping absolute magnitude range, the two measured LFs are in good agreement within the $1-\sigma$ errors.

\citet{saracco2006}  estimated the $J$-band LF in three redshift ranges, namely $z<0.8$, $0.8<z<1.9$ and $1.9<z<4.0$, using 101, 100, and 84 galaxies, respectively, collected from the HDF-S data and complemented by VLT-ISAAC J,H and K imaging. The LFs from \citet{saracco2006} are compared to our measurements in Figure~\ref{fig:comp_saracco}. At  $z\approx 1.5$, the absolute magnitude ranges of the two determinations  are quite different. Their lack of points at the bright end, presumably due to the very small field of the HDF-S compared to ours, is compensated by a deeper limit at the faint end. Their Schechter representation of the LF is flatter ($\alpha=-0.94$) than our measurement of the LF at $1.5<z<2.0$, and it presents a dimmer characteristic magnitude (by $\sim$1~mag). However, when directly comparing the $1/V_{\rm max}$ estimate from \citet{saracco2006} with our non parametric  $1.5<z<2.0$ LF, we find good agreement (see Figure \ref{fig:comp_saracco}, left panel) with measurements lying within the 1~$\sigma$ error bars. Despite these differences, our Schechter parameterization is substantially compatible also with their  points. We would like to note that our composite catalogue provides an improved sampling of the bright end, resulting in overall better constrained Schechter parameters, due to the strong correlation among these parameters. In the highest redshift bin, the differences in the Schechter parameters are still present, with the faint-end slope $\alpha$ being the parameter showing the largest difference. As for the lower-redshift bin, when comparing our non parametric estimate with their points, we find a good agreement (see the right panel in Figure \ref{fig:comp_saracco}).

\begin{figure}
\includegraphics[width=8.5cm]{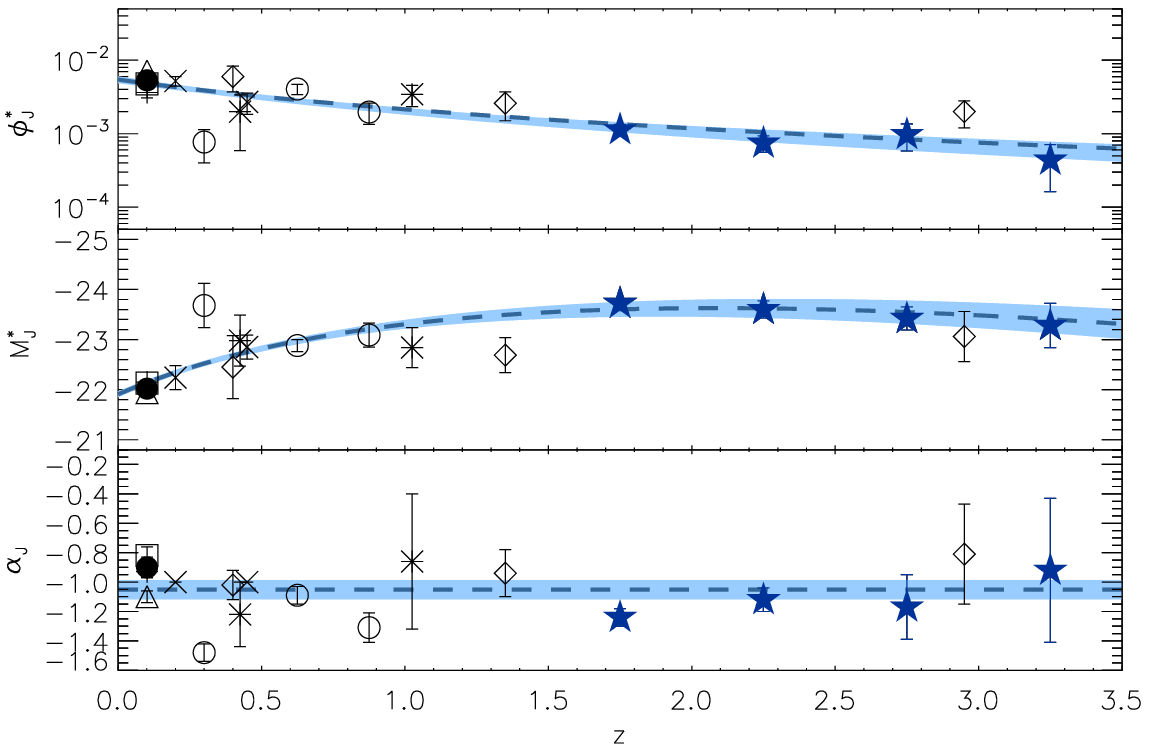}
\caption{Evolution of the Schechter parameters of the J LF as a function of redshift. Estimates from this work are shown as filled blue stars. Measurements from the literature are also plotted (vertical crosses: \citealt{cole2001}; asterisks: \citealt{pozzetti2003}; crosses: \citealt{feulner2003}; open squares: \citealt{eke2005}; open circles: \citealt{dahlen2005}; open diamonds: \citealt{saracco2006}; open triangles: \citealt{jones2006}; filled circle: \citealt{hill2010}). Top panel shows the data for $\phi^*$. The dashed line represents Eq. \ref{eq:fit_phi} (see text for details); the middle panel presents the measurements for $M^*$, with the dashed line representing Eq. \ref{eq:sch_ms}; in the lower panel the faint end slopes $\alpha$ together with the average value (dashed line) are shown. The filled regions correspond to $95\%$ confidence level intervals for the resulting fitting curves.}
\label{fig:sch_evol_J}
\end{figure}

\subsection{Discussion}

In Figure \ref{fig:sch_evol_J} we compare the Schechter parameters  for the rest-frame $J$ band obtained in this work with those available in the literature as a function of redshift.

The upper panel shows the evolution of $\phi^*$ as a function of redshift, showing $\phi^*$ monotonically decreasing with increasing $z$. By $z\simeq 2$, $\phi^*$ has decreased by approximately an order of magnitude compared to the local values. The following parameterization was adopted to model the observed evolution of $\phi^*$ with $z$:
\begin{equation}
\phi^*(z)= \theta \exp{\left[\gamma/(1+z)^\beta\right]}
\label{eq:fit_phi}
\end{equation}
where $\theta$, $\gamma$, and $\beta$ are the free parameters. The best-fit values obtained for the parameters of the rest-frame $J$ band are  $\theta_J= 2.6 \pm 0.9 \times 10^{-6}$ mag$^{-1}$ Mpc$^{-3}$, $\gamma_J=7.7 \pm 0.3$, $\beta_J=0.22$. The $\beta$ parameter was estimated together with the other two in the first instance of the best fitting procedure, and kept fix in a second iteration. The quoted errors refer to the second iteration. Equation \ref{eq:fit_phi} is plotted as a dashed line in the upper panel of Figure \ref{fig:sch_evol_J}.

In the middle panel of Fig. \ref{fig:sch_evol_J} we present the evolution of $M^*$ as a function of redshift. The data show a brightening of $M^*$ from the local universe to $z\simeq2$, followed by a slow dimming. In analogy to the LF shape by \citet{schechter1976}, it is then possible to introduce the following ad-hoc representation for $M^*(z)$:
\begin{equation}
M^*(z)=\mu \left[ (1+z)/(1+z^*) \right]^\eta \exp{\left[-(1+z)/(1+z^*)\right]}
\label{eq:sch_ms}
\end{equation}
with $\mu$, $z^*$, and $\eta$ free parameters to be determined. By performing a least-square fit to the available data we obtain the following values: $\mu_J=-37.6\pm 1.7$ mag, $z^*_J=17.5\pm 4.6$, $\eta_J=0.16\pm 0.02$. The resulting curve is plotted as a dashed line in the middle panel of Figure \ref{fig:sch_evol_J}.

The lower panel illustrates the behavior of $\alpha$ as a function of redshift. The error bars are here generally large and do not allow to properly evaluate the presence of evolution as a function of z. Therefore we limit ourselves to compute an average value, resulting in $\bar{\alpha}=-1.05 \pm 0.03$.

\begin{figure}
\includegraphics[width=8.5cm]{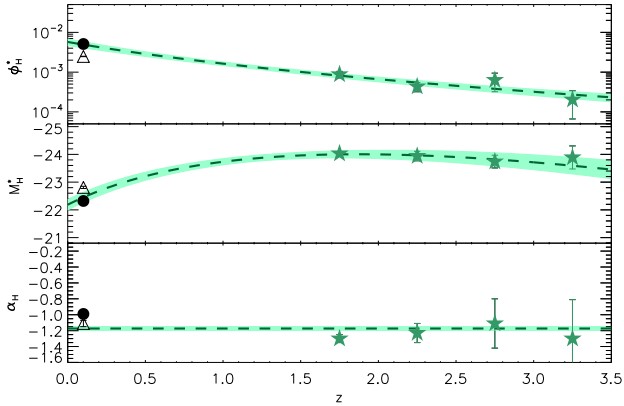}
\caption{Evolution of the Schechter $\phi^*$, $M^*$ and $\alpha$ parameters from the H LF as a function of redshift. See caption to Figure \ref{fig:sch_evol_J} for details.}
\label{fig:sch_evol_H}
\end{figure}

Figure \ref{fig:sch_evol_H} shows the plots of the Schechter parameters as a function of redshift corresponding to the rest-frame $H$ band. For this band, there are only two determinations of the LF from the literature, making it more challenging to determine the evolution with redshift. Despite this, we applied the same analysis done for the rest-frame $J$ band, obtaining $\theta_H=2.0 \pm 1.3 \times 10^{-6} $ mag$^{-1}$ Mpc$^{-3}$, $\gamma_H=7.8 \pm 0.5$, $\beta_H=0.30$ for the parameters of Eq. \ref{eq:fit_phi};  $\mu_H=-40.0\pm 9.3$, $z^*_H=13.9\pm 6.3$, $\eta_H=0.19\pm 0.06$ for Eq. \ref{eq:sch_ms} and $\bar{\alpha}=-1.15 \pm 0.02$. The resulting best-fits are shown as dashed curves in Figure \ref{fig:sch_evol_H}.

\subsection{Luminosity densities}

\begin{figure}
\includegraphics[width=8.5cm]{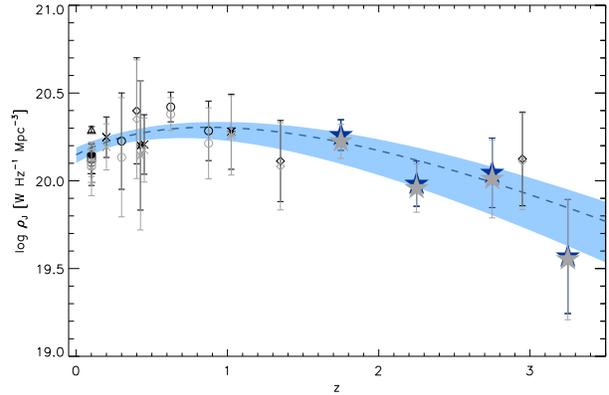}
\caption{Luminosity density $\rho_J$ as computed from our Schechter parameters (filled blue stars) and compared with the available data in the redshift range [0,3.0]. Grayed symbols indicate the luminosity density computed assuming an absolute magnitude limit of $M_J=-20.0$. The dashed line represents the LD obtained directly from Eq. \ref{eq:rho} in terms of Eq. \ref{eq:fit_phi} and \ref{eq:sch_ms}, with $95\%$ confidence level indicated by the filled region. Plotting symbols same as for Fig. \ref{fig:sch_evol_J}.}
\label{fig:lumden_J}
\end{figure}

\begin{figure}
\includegraphics[width=8.5cm]{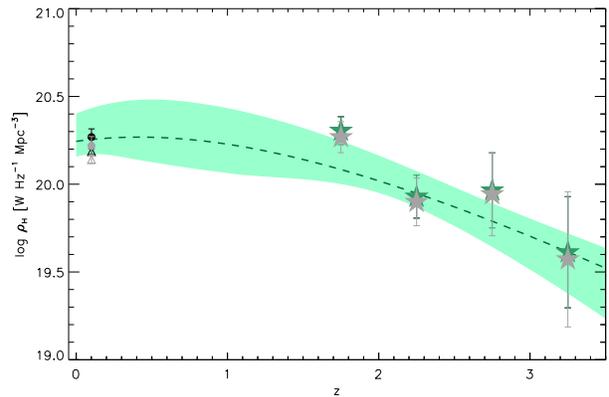}
\caption{Luminosity density $\rho_H$ as computed from our Schechter parameters (filled black stars) and compared with the available data. See caption to Fig. \ref{fig:lumden_J} for details. }
\label{fig:lumden_H}
\end{figure}

Here we present our measurements of the luminosity density (LD). Given the coupling between the Schechter parameters $\alpha$ and $M^*$, the luminosity density is a robust tool to characterize the evolution of the LF with cosmic time. The LD was obtained in the standard way, i.e. as:
 
 \begin{equation}
 \rho_J=\int_0^{+\infty}L \Phi(L) dL = \Gamma(2+\alpha) L^* \phi^*
 \label{eq:rho}
 \end{equation}
where the last equality holds when assuming  a Schechter parametrization for the $\Phi(L)$. This means that we are assuming that the Schechter distribution is a good representation of the underlying luminosity function. Figure \ref{fig:lumden_J} shows the evolution of the luminosity density in the $J$ filter, while Figure \ref{fig:lumden_H} displays the corresponding plot for the rest-frame $H$ band. Values of the luminosity density at each redshift and for each filter are presented in Table \ref{tab:lumden}.

\begin{table}
\begin{tabular}{ccccc}
Filter & $z$ range & log $\rho$ & log $\rho^*$  & log $\bar{\rho}$\\
\hline
$J$ &1.5-2.0  & $20.26 \pm  0.08$ & $20.22 \pm 0.09$ & $20.22 \pm 0.09$ \\ 
& 2.0-2.5 & $19.98 \pm  0.12$ & $19.96 \pm 0.13$   & $19.93 \pm 0.13$ \\
&2.5-3.0 & $20.05 \pm  0.18$ & $20.01 \pm 0.20$   & $19.93 \pm 0.24$ \\
& 3.0-3.5 & $19.57 \pm  0.30$ & $19.55 \pm 0.32$  & $19.45 \pm 0.39$ \\
&&&\\
$H$ &1.5-2.0  & $20.30  \pm  0.08$ & $20.27 \pm 0.09$ & $20.27 \pm 0.09$\\ 
& 2.0-2.5 & $19.93 \pm 0.12$ &  $19.90 \pm  0.14$   & $19.87 \pm 0.14$ \\
&2.5-3.0 &  $19.96 \pm 0.21$ &  $19.94 \pm  0.24$  & $19.85 \pm 0.28$ \\
& 3.0-3.5 &  $19.61   \pm 0.32$ &  $19.57 \pm  0.38$ & $19.39 \pm 0.52$ \\
\end{tabular}
\caption{Luminosity density in logarithmic scale and expressed in units of log[W Hz$^{-1}$ Mpc$^{-3}$], in the rest-frame $J$ and $H$ bands. Quoted errors include the effects of cosmic variance, which is the dominant source of random uncertainties. In the third column we report the luminosity density computed using Eq. \ref{eq:rho} and corresponding to the black stars in Figures \ref{fig:lumden_J} and \ref{fig:lumden_H}; the values of the luminosity density in the fourth column  (log $\rho^*$) reflect the upper limit in absolute magnitude corresponding to $M_{lim}=-20$, which we imposed in order to limit the effect of the uncertainties in the determination of $\alpha$ (grey stars in the same figures). The values in the last column (log $\bar{\rho}$) list the luminosity densities computed using the corresponding absolute magnitude limits at each redshift bin and for each band, i.e., $M_{J,lim}=-20, -21,-21.5,-22$, and $M_{H,lim}=-20,-21,-22,-22.5$ for the redshift intervals centered at $z=1.75, 2.25, 2.75, 3.25$, respectively. }
\label{tab:lumden}
\end{table}

In order to be less sensitive to the derived faint end slope of the LF, we also computed the luminosity density assuming two different limiting absolute magnitudes. First, the luminosity density $\bar{\rho}$ was derived using the absolute magnitude limits of our survey in each redshift bin, i.e., $M_{J,lim}=-20, -21,-21.5,-22$, and $M_{H,lim}=-20,-21,-22,-22.5$, for the redshift intervals centered at $z=1.75, 2.25, 2.75, 3.25$, respectively. Second, the luminosity density $\rho^*$ was derived assuming a limiting absolute magnitude equal to the brightest limit over the entire targeted redshift range, i.e. $M_{lim}=-20.0$. The values of $\rho^*$ are also plotted in Figures~\ref{fig:lumden_J} and \ref{fig:lumden_H} as grey symbols.

\begin{figure*}
\includegraphics[width=18.2cm]{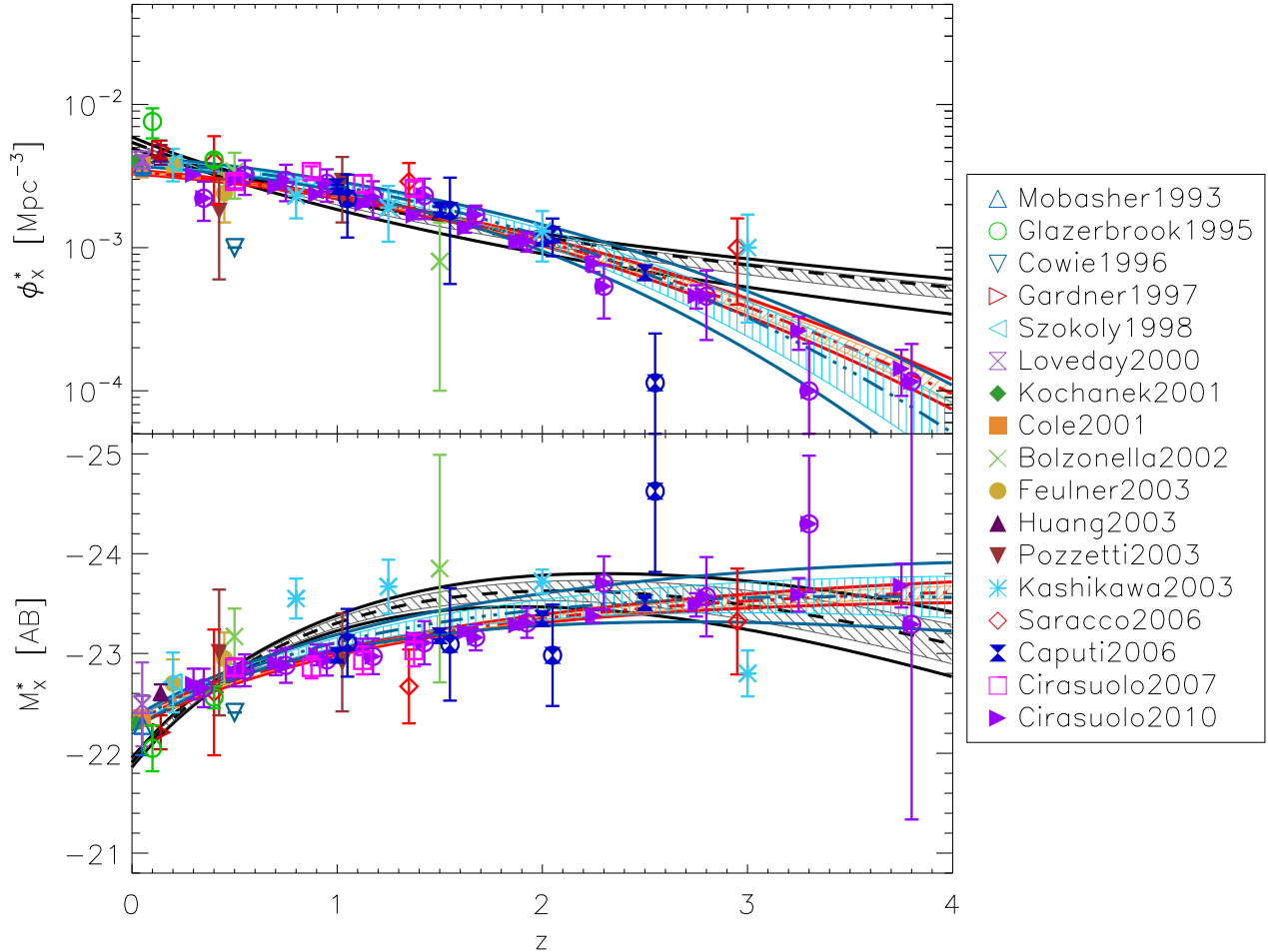}
\caption{Top panel: comparison between the evolution with redshift of the Schechter parameter $\phi^*$ in the rest-frame $J$ and $K$ (coloured points - see the legend on the right for details) bands. For the \citet{caputi2006} and \citet{cirasuolo2010} LFs, the parameter values recovered from a Schechter fit to the $1/V_{\rm max}$ LFs are identified by an additional circle and are displaced by an arbitrary amount of 0.05 in redshift to increase readability. The black dashed curve shows the best-fit parameterizations obtained for the rest-frame $J$ band. The red and the blue curves mark the best-fit parameterization for the $K$ band using the original ML from \citet{caputi2006} and \citet{cirasuolo2010} and that using the points from the Schechter fit to their $1/V_{\rm max}$ LFs, respectively. The hatched regions delimit the $68\%$ confidence level for the fit parameters for the rest-frame $J$ band (grey), $K$ band (orange), and $K$ band with Schechter parameters obtained from fitting the $1/V_{\rm max}$ LFs (cyan), respectively. 95\% confidence intervals are also plotted as coloured solid black, red, and blue curves, respectively. Bottom panel: evolution with redshift of the rest-frame $J$- and $K$-band characteristic magnitude $M^*$. Same plotting conventions as top panel apply.}
\label{fig:LF_K_comp}
\end{figure*}

\begin{figure*}
\includegraphics[width=18.2cm]{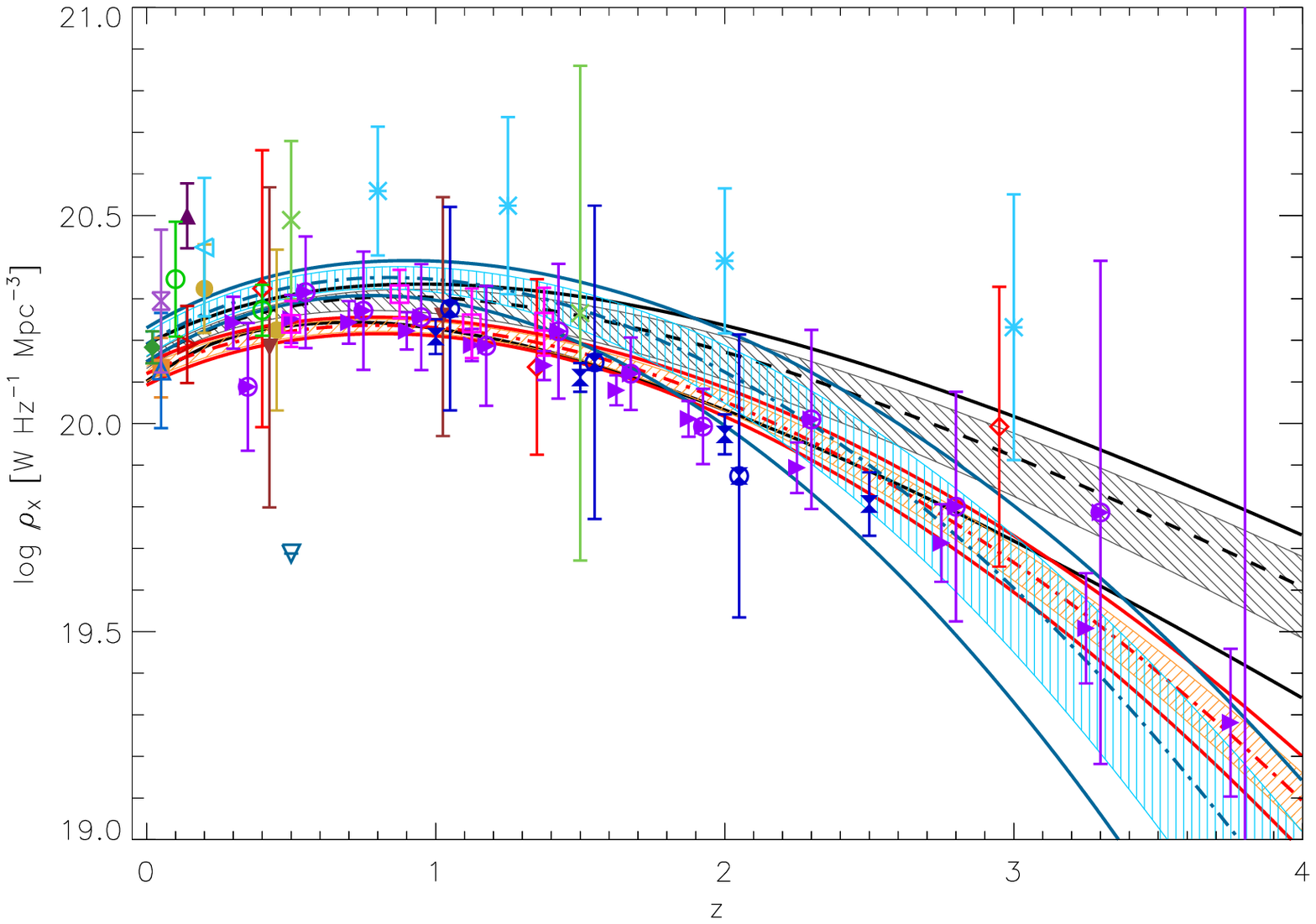}
\caption{Comparison between the evolution with redshift of the LD in the rest-frame $J$ (black curves and gray hatched regions - same plotting conventions as for Fig. \ref{fig:sch_evol_J}) and $K$ bands (coloured points - same plotting conventions as for Fig. \ref{fig:LF_K_comp}). The hatched regions delimit the $68\%$ confidence levels for the fit parameters, while the external coloured solid curves delimit the 95\% confidence levels.}
\label{fig:LD_K_comp}
\end{figure*}

The overall plot of the $J$-band luminosity density shows a constant value for $z \lesssim 0.8-1.0$. At $z\approx 0.8-1.0$, the luminosity density starts to decrease down to $z\approx 3.5$, where $\rho_J$ is 16\% of the $z=0$ value. This can be better visualized by comparing this plot with the top and middle panels of Figure \ref{fig:sch_evol_J}. Here we see in fact that for $z\lesssim1$ the decrease in number of galaxies is balanced by a brightening of the characteristic magnitude. At $z \gtrsim 1$, both quantities decrease, with the resulting decrease of the luminosity density.

Using the expression of Eq. \ref{eq:fit_phi} and Eq. \ref{eq:sch_ms} in Eq. \ref{eq:rho}, it is possible to obtain a functional representation of the luminosity density. The dashed line in Figure \ref{fig:lumden_J} represents the luminosity density for the rest-frame $J$ band obtained with this method and adopting the best-fit values of the parameters previously recovered. The agreement with the points is good over the entire redshift range. We would like to stress that no best fit has been done using the data of the luminosity density.

Similarly to the case of the LF, the luminosity density in the $H$ filter has been poorly studied, so that it is more difficult to derive its evolution. Our data however indicate a decline with redshift of the LD, similar in shape to the one found in the $J$ band, with a faster evolution from $z=3.5$ to $z=1.5$, followed by a much slower evolution, decreasing by a factor of $\sim 7$ from $z=0$ to $z=3.5$.  An exercise similar to what done for the $J$-band luminosity density, introducing our parameterizations, is shown as a dashed line in Figure \ref{fig:lumden_H}. The agreement is quite good over the whole redshift range, although more measurements are necessary at intermediate redshifts ($z<1.5$).

\subsection{Comparison with rest-frame $K$-band LF and LD}

In Fig.~\ref{fig:LF_K_comp} we present the evolution of the Schechter parameters $\phi^*$ (top panel) and $M^*$ (bottom panel) for the rest-frame $K$ band, collected from the literature (coloured points - see legend for details) and overplotted to the corresponding parameterization of the rest-frame $J$ band (taken from Fig. \ref{fig:sch_evol_J}; black line). The previously measured rest-frame $K$-band LFs are taken from \citet{mobasher93,glazebrook1995,cowie96,gardner97,szokoly98,loveday2000,kochanek2001,cole2001,bolzonella2002,feulner2003,huang2003,pozzetti2003,kashikawa2003,saracco2006,caputi2006,cirasuolo2007,cirasuolo2010} - see also Table 5 in \citet{saracco2006} which we used as reference for the literature works. 

Using the rest-frame $K$-band data, we performed the same analysis as done for the rest-frame $J$ and $H$ bands, modeling the evolution of the Schechter parameters with redshift using Eq.~\ref{eq:fit_phi} and \ref{eq:sch_ms}. The best-fit values of the parameters obtained in modeling of the $K$-band points are: $\theta_K=3.8 \pm 0.1 \times 10^{-3} $ mag$^{-1}$ Mpc$^{-3}$, $\gamma_K=-0.11 \pm 0.02$, $\beta_K=-2.2 \pm 0.1$ for the parameters of Eq. \ref{eq:fit_phi};  $\mu_K=-29.6\pm 0.4$, $z^*_K=113\pm 46$, $\eta_K=0.058\pm 0.007$ for Eq. \ref{eq:sch_ms}; and $\bar{\alpha}_K=-1.12 \pm 0.16$. The resulting curves are plotted in Fig. \ref{fig:LF_K_comp} as red dot-dashed curves, together with those already discussed for the rest-frame $J$ band (black dashed curves).

The rest-frame $K$-band characteristic density, $\phi_{\rm K}^*$, decreases by 
a factor of $\sim 15$ from $z\sim 0$ to $z\sim 3.3$, about twice as much than 
the redshift evolution of the rest-frame $J$-band characteristic density, 
$\phi_{\rm J}^*$. Specifically, the data in the rest-frame $J$ band indicate an 
evolution with redshift broadly consistent to the rest-frame $K$ band out to 
$z\sim2.3$, and a milder evolution at $z\gtrsim 2$. Quantitatively, while 
$\phi_{\rm J}^*$ decreases by a factor of $\sim 2$ from $z\sim1.5$ to 
$z\sim3.3$, $\phi_{\rm K}^*$ decreases by a factor of $\sim 5$ over the same 
redshift interval, although these differences are significant only at the 
2~$\sigma$ level.

Differences between the rest-frame $J$ band and the rest-frame $K$ band are 
also found when comparing the evolution with redshift of the characteristic 
magnitudes $M_{\rm K}^*$ and $M_{\rm J}^*$ (bottom panel of 
Fig.~\ref{fig:LF_K_comp}). At $z \lesssim 1$, similar evolutions with redshift 
of $M_{\rm K}^*$ and $M_{\rm J}^*$ are found, with the characteristic magnitudes 
brightening by $\sim 0.8$~mag from $z\sim0$ to $z\sim1$. However, for 
$z \gtrsim 1$, $M^*_K$ is monotonically decreasing,  brightening by 0.5 mag in 
the range $z\in[1.5, 3.3]$, whereas $M^*_J$ shows a small dimming (if any) 
over the same redshift interval, after reaching its brightest value somewhere 
in the redshift interval $1.5<z<2.5$. We note that, also in the case of $M^*$, 
these differences are only marginally significant (within 2~$\sigma$).

Figure~\ref{fig:LD_K_comp} shows the comparison of the evolutions with 
redshift of the rest-frame $K$- and $J$-band LDs. As the differences in 
the evolutions with redshift of $\phi^*$ and $M^*$ between the $J$ and 
$K$ bands go in opposite directions, in the computation of the LDs these 
differences partly cancel out. As shown in Figure~\ref{fig:LD_K_comp}, the 
evolution of the rest-frame $K$-band LD with redshift is qualitatively similar 
to the evolution with redshift of the rest-frame $J$-band LD, decreasing by a 
factor of almost $\sim2$ from $z\sim 0$ to $z\sim 2.5$.

The evolution of the $K$-band LF (and, consequently, LD) at $z\gtrsim1$ is 
dominated by the measurements by \citet{caputi2006} and \citet{cirasuolo2010}, 
obtained adopting ad-hoc parameterizations for $\phi^{*}_{K}$ and $M^{*}_{K}$, 
with $\phi^*_K (z)\propto \exp(-z^{\eta_\phi})$ and $M^*_K(z) \propto z^{\eta_M}$ (see Eq. 2 and 3 in \citealt{cirasuolo2010} for details). In order to 
verify that the derived redshift evolution in the rest-frame $K$ band does not 
depend on the adopted parameterizations, we computed the Schechter function 
parameters by fitting their $1/V_{\rm max}$ estimates with a Schechter function 
with $M^*_K,\phi^*_K$ and $\alpha_K$ as free parameters. The results are 
indicated in Fig. \ref{fig:LF_K_comp} and Fig. \ref{fig:LD_K_comp} by an 
additional circle around the corresponding symbol, while our 
evolutionary-model best fits (Eq.  \ref{eq:fit_phi} and \ref{eq:sch_ms}) are 
shown by the blue curves. The values of the Schechter parameters recovered 
from the fit to the $1/V_{\rm max}$ measurements are in broad agreement with 
those obtained from the ML analysis in combination with the adopted 
$\phi^{*}_{K}(z)$ and $M^{*}_{K}(z)$ parameterizations, although with larger 
error bars. Consequently, the evolution with redshift of $\phi^{*}_{K}$ and 
$M^{*}_{K}$ inferred from the Schechter parameters obtained from the 
fit to the $1/V_{\rm max}$ measurements partly overlap to that obtained using 
the original ML data of \citet{caputi2006} and \citet{cirasuolo2010}, although 
with significantly larger errors.

The final picture is that globally the evolutions of the rest-frame $J-$ 
and $K-$band LFs are (marginally) different within a confidence level not 
exceeding the 95\% (i.e. roughly 2 $\sigma$). The LDs in the two rest-frame 
bands are consistent up to $z\sim2.5$. At $z \gtrsim 2.5$, there seems 
evidence of a faster decrease with redshift of the rest-frame $K$-band LD 
compared to the rest-frame $J$-band LD at 68\% confidence level. However, 
the evolutions of the LDs in the two bands are comparable when considering 
the 95\% confidence level. We note that the measurements of the $K$-band LDs 
derived from the Schechter function fits to the $1/V_{\rm max}$ analysis from 
\citet{caputi2006} and \citet{cirasuolo2010} are consistent with our estimated 
rest-frame $J$-band LDs within the errors. More accurate measurements of the 
LFs at $z\gtrsim 2.5$ in both the $J$ and $K$ bands are required to further 
investigate possible differences in the redshift evolution of the LDs in the 
two NIR bands.

Figures~\ref{fig:LF_K_comp} and \ref{fig:LD_K_comp} show the presence of 
significant scatter in the measurements of the rest-frame $K$-band 
characteristic magnitudes $M_{\rm K}^*$ and LDs from different works (up to a 
factor of $\sim$2-3 when comparing the works of \citealt{kashikawa2003} and 
\citealt{cirasuolo2010}). Whereas it is difficult to assess the sources of such 
scatter and beyond the scopes of this work, we note that potential sources 
for such a significant scatter are twofold. 

First, any direct measurements of the rest-frame $K$-band LF and LD at 
$z \gtrsim 0.6$ require observations at wavelengths longer than 2.2~$\mu$m, 
provided by, e.g., Spitzer-IRAC. As \citet{caputi2006}, \citet{cirasuolo2007}, 
and \citet{cirasuolo2010} are the only previous works targeting the rest-frame 
$K$ band including IRAC data in their analysis, all the other works do not 
directly probe the rest-frame $K$ band, but instead rely on stellar population 
synthesis models or empirical templates to extrapolate the rest-frame $K$-band 
magnitudes of galaxies. Therefore, some of the scatter observed in the 
measurements of the rest-frame $K$-band LF parameters and LDs at 
$z \lesssim 2.5$ could be due to differences in the adopted stellar population 
synthesis models, such as differences in the implementation of the TP-AGB 
phase (i.e., \citealt{bruzual2003}; \citealt{maraston2005}). Note that 
observations at wavelengths longer than provided by IRAC would be needed to 
directly probe the rest-frame $K$ band at $z \gtrsim 2.6$, so all measurements 
of the rest-frame $K$-band LF and LD at $z \gtrsim 2.6$ rely on stellar models. 

Second, emission associated to obscured (e.g., type-II) AGN (e.g., 
continuum emission from the dusty torus) can be significant in the rest-frame 
$K$ band and at longer wavelengths. At low redshifts ($z<0.15$), nuclear 
contamination in the $K$ band from (obscured) AGN has been shown to be 
$\sim$25\% on average in radio galaxies, and as high as 40\% (on average) in 
broad-lined radio galaxies \citep{inskip2010}. AGN contamination are generally 
much smaller in the rest-frame $J$ or $H$ bands, with an average contamination 
of $\sim$3-4\% (e.g., \citealt{floyd2008}). We note that, while these levels 
of AGN contamination have been derived for radio-loud AGNs, which represent 
only $\sim$10\% of the overall AGN population, radio-quiet AGNs are expected 
to be characterized by similar levels of contamination, as a result of the 
similar orientation-based unification scheme \citep{antonucci1993,honig2006}. Moreover, 
evidence for an increasing fraction of AGN as a function of stellar mass at 
$z>2$ has been recently found, with an AGN fraction as high as 70\% for 
massive galaxies at $2.0<z<2.7$ \citep{kriek2007}, and potentially even higher 
AGN fractions in massive galaxies at $3<z<4$ \citep{marchesini2010}. As the 
rest-frame $K$ band is directly probed by IRAC 3.6, 4.5, 5.8, and 8~$\mu$m 
channels at $z\sim0.6$, 1, 1.6, and 2.6, respectively, the AGN-associated 
emission can potentially contaminate the measurements of the rest-frame 
$K$-band LFs and LDs, especially at the bright end, potentially contributing 
to the observed scatter in the rest-frame $K$-band measurements of $M^*$ and 
LD. An accurate quantification of the contamination from AGN to the rest-frame 
K band would require a detailed modeling of the SEDs from the X-ray to the 
infrared associated to each object (see, e.g., \citealt{lusso2012}), which is 
beyond the scope of this work.

\section{Conclusions}
In the present work, we used a composite sample constructed from deep multi-wavelength publicly available photometric catalogues from the MUSYC, FIRES and FIREWORKS survey. The availability of \emph{Spitzer}-IRAC data in the 3.6, 4.5, 5.8, and 8$\mu$m channels allows us to robustly estimate the LFs and LDs in the rest-frame $J$  and $H$ bands with a minimum dependence on the SED templates up to $z=3.5$. Ours is the first measurement of the rest-frame $H$-band LF at $z>0$ to-date. We determined the LF with three independent methods, namely the $1/V_{\rm max}$, the SWML ,and the STYML methods, finding that they agree well with each other. Uncertainties introduced by the cosmic variance were estimated using two distinct methods, one by \citet{moster2010} and the other by \citet{driver2010}. We find that, for our data, the two approaches broadly agree.

Our rest-frame $J$-band LF is consistent with previous determination by \citet{saracco2006}, although the recovered Schechter parameters $M^*$ and $\alpha$ are consistent only at the 2$\sigma$ level. This might be due to the limited range in rest-frame magnitudes probed by the sample in \citet{saracco2006} and by large errors due to cosmic variance given their small surveyed area. Our $J$-band LF is consistent also with the LF measured by \citet{pozzetti2003} at $z\sim1.5$.

We analyzed the evolution with redshift of the Schechter function parameters, making full use of the data available from the literature. We found that the faint end slope $\alpha$ of the LF is nearly constant over the whole redshift range, with $\alpha_J=-1.05\pm0.03$ and $\alpha_H=-1.15\pm0.02$ in the $J$ and $H$ bands, respectively. The characteristic density $\phi^*$ decreases by a factor of $\approx 6$ from $z\sim0$ to $z=1.75$, and by a factor of $\approx 3$ from $z\sim 1.75$ to $z=3.25$. We introduced a parameterization based on an exponential form for the evolution of $\phi^*$ as a function of $z$. The fit of this function to the available data shows good agreement, especially for the rest-frame $J$ band, where more data from the literature are available in the redshift range $z \in[0, 1]$, complementing our measurements at $z>1.5$.

The characteristic magnitude $M^*$ is found to brighten from $z=0$ to $z\sim2$ by $\sim0.8$~mag, whereas $M^*$ gets fainter with increasing redshift at $z \gtrsim 2$. We adopted a \cite{schechter1976}-like expression for its description, resulting in a good representation of the observed evolution.

We computed the LD in the rest-frame $J$ and $H$ bands, using the Schechter parameters previously determined. The LD is nearly constant
up to $z\approx 1$ and decreases as a power-law by a factor of $\approx 6$ from $z\approx1$ to $z=3.25$.

We compared the evolution with redshift of the LFs and LDs in the 
rest-frame $J$ and $K$ bands. The Schechter parameters $\phi^*$ and $M^*$ in 
the rest-frame $K$ band show different evolutions with redshift with respect 
to the $J$ band, although these differences are only marginally significant at 
the 2-$\sigma$ level. Specifically, the decrease of the characteristic density 
with redshift appears faster in the $K$ than in the $J$ band at 
$z \gtrsim 1.5$. The characteristic magnitude in the rest-frame $K$ band 
brightens with redshift over the whole redshift interval $0<z<4$, whereas 
$M^*_{\rm J}$ gets brighter from $z=0$ to $z\sim2$, and then slowly gets 
fainter out to $z=3.5$ (although a constant value of $M^*_{\rm J}$ at $z>1.5$ 
is consistent within the uncertainties). Most of these differences cancel out 
when computing the LDs, with similar evolutions with redshift of the 
rest-frame $J$- and $K$-band LDs out to $z\sim2.5$. Evidence for a faster 
decrease with increasing redshift of the rest-frame $K$-band LD at 
$z\gtrsim2.5$ seems present. However, these differences are significant only 
to a 95\% level (2 $\sigma$). Large errors at $z>2$ in both the $J$ and $K$ 
bands prevent to firmly assess differences between the evolution with 
redshift of the $J$- and $K$-band LFs and LDs.

In order to further constrain the rest-frame $J$- and $H$-band LFs, a larger area is needed to better probe the bright end and reduce the impact of field-to-field variations and low number statistics. Better photometric redshift estimates are also needed to improve the LF measurements at the bright end, by, e.g., reducing the impact of catastrophic outliers. The recently publicly released NEWFIRM Medium-Band Survey \citep{whitaker2011} will provide the dataset to significantly improve on all of these aspects. Ongoing very deep ground-based surveys, such as the Ultra-VISTA survey\footnote{http://www.eso.org/sci/observing/policies/PublicSurveys/scienePublicSurveys.html}, and space-based surveys with WFC3 on the \emph{Hubble Space Telescope}, such as the CANDELS \citep{grogin2011,koekemoer2011} and 3D-HST (Brammer et al. 2011, in prep.) surveys, will allow for much improved constraints of the faint end of the rest-frame NIR LF and of the contribution of low-luminosity galaxies to the total NIR LD. 

\section*{Acknowledgments}
We thank all the members of the FIRES, FIREWORKS, and MUSYC collaborations for their contribution to this research. We thank A. Fernandez-Soto and P. Saracco for useful comments and constructive discussions. MUSYC has greatly benefited from the support of Fundacion Andes and the Yale Astronomy Department. This work is based on observations with the Spitzer Space Telescope, which is operated by the Jet Propulsion Laboratory (JPL), California Institute of Technology under NASA contract 1407; based on observations with the NASA/ESA Hubble Space Telescope, obtained at the Space Telescope Science Institute, which is operated by AURA, Inc., under NASA contract NAS5-26555; based on observations collected at the European Southern Observatories, Chile (ESO Programme LP164.O-0612, 168.A-0485, 170.A-0788, 074.A-0709, 275.A-5060, and 171.A-3045); based on observations obtained at the Cerro Tololo Inter-American Observatory, a division of the National Optical Astronomy Observatories, which is operated by the Association of Universities for Research in Astronomy, Inc., under cooperative agreement with the National Science Foundation.

\end{document}